\documentclass[11pt]{article}
\usepackage{amsmath,amssymb,multirow,jheppub,graphicx}
\usepackage{verbatim}   
\usepackage[all]{xy}
\usepackage{bm}  
\usepackage{slashed}
\usepackage{color}
\usepackage{eufrak}

\allowdisplaybreaks[1]

\def\Z{\mathbb{Z}} 
\def\R{\mathbb{R}}

\def\Im{\text{Im}}
\def\Re{\text{Re}}

\def\tr{\text{tr}}

\def\hat{\widehat}

\def\N{{\mathcal N}}

\def\n{{\nu}}
\def\x{{\xi}}

\def\O{{\Omega}}

\def\half{\coeff 12}

\def\N{{\cal N}}
\def\T{{\cal T}}

\def\I{{\cal I}} 

\def\S_1{{\widetilde {S_1}}}

\def\R{{\mathbb R}}

\def\tr{{\rm tr}}
\def\Re{{\rm Re}\,}
\def\x{\mathbf x}

\def\Z{{\mathbb Z}}

\def\Dslash{{\rlap{\raise 1pt \hbox{$\>/$}}D}}
\def\O{{\cal O}}

\def\Im{\mathrm {Im}}

\def\half{\textstyle \frac{1}{2}}
\def\OI{ \overline {\cal I}}
\def\n{ \overline n}

\title{\huge Theta  dependence,  sign problems and 
topological interference}

\author[1,2]{ Mithat \"Unsal}
\affiliation[1]{Department of Physics and Astronomy, SFSU, San Francisco, CA 94132, USA }
   \affiliation[2]{SLAC and Physics Department, Stanford University, Stanford, CA 94025/94305, USA}
\emailAdd{unsal.mithat@gmail.com}

\abstract{
In a Euclidean path integral formulation of gauge theory and quantum mechanics,    the $\theta$-term induces a sign problem, and relatedly, 
 a complex  phase for the fugacity of topological defects;  whereas  in Minkowskian  formulation,  
 it induces a   topological (geometric)  phase multiplying ordinary path-amplitudes. 
    In an  $SU(2)$   Yang-Mills  theory which admits a semi-classical limit, 
     we show that the     complex fugacity  
generates    interference between  Euclidean path histories, {\it i.e.},  
monopole-instanton events, and    radically alters the vacuum structure. 
 At  $\theta=0$, a mass gap  is due to the  monopole-instanton plasma, and the theory has a unique vacuum. 
 At $\theta=\pi$, the monopole induced mass gap  vanishes, despite the fact that monopole density is independent  of $\theta$, due to destructive topological interference.  The theory has two options: to remain gapless or to be gapped with a two-fold degenerate vacua. We show the latter is realized by the  magnetic bion mechanism, and the two-vacua are realization of spontaneous CP-breaking.   
 
The effect of the $\theta$-term  in the circle-compactified  gauge  theory is a generalization of  Aharonov-Bohm effect,  and  the  geometric (Berry) phase. 
 As $\theta$ varies from $0$ to $\pi$,   the gauge theory interpolates between even- and odd-integer  spin quantum anti-ferromagnets on   two spatial dimensional bi-partite lattices, which have ground state degeneracies one  and two, respectively,  as it is  in gauge theory at $\theta=0$ and  $\theta=\pi$. 
}

\begin{document}
\maketitle
\section{\label{sec:level1}Topological terms}
Topological terms in quantum field theories,  for example $\theta $, Chern-Simons, and WZW, 
may affect the low energy theory in non-trivial ways. They also render Euclidean action complex, and introduce a sign problem in numerical simulations based on the Euclidean path integral formulations.   
Questions about the dependence of the mass gap and the spectrum  on the $\theta$ angle in Yang-Mills theory are  physical, but also out of reach  due to strong coupling.  A way to gain  insight into a strongly coupled and asymptotically free gauge  theory  is to move to  a simpler theory which resembles the target theory as much as possible\footnote{We  demand that  the simpler theory should be  asymptotically  free,   should possess the same global symmetries, and
identical  matter content (for light or massless fields) as the original theory. If possible, it should also be continuously connected to the original theory, so that  maximum amount of data can be extracted about the original theory.},
and  which shares the same universality properties as the original theory.

  In this work,  we  report on a small step on  $\theta$-angle dependence  of  observables in 
$SU(2)$ Yang-Mills theory by using continuity, and  deformed Yang-Mills theory \cite{Unsal:2008ch, Shifman:2008ja}. The deformed theory, on small  $\R^3 \times S^1$, is continuously connected to the pure Yang-Mills theory  on large $\R^3 \times S^1$ and  $\R^4$   in the sense that  the only global symmetry of the compactified theory, the center symmetry, is unbroken in both regimes. 
 Using this framework, 
we calculate the vacuum energy density,  mass gap,   string tension, deconfinement temperature, and CP-realization    by using  semi-classical field theory at decidedly small values of the number of colors $N$,  and for all values of  $\theta \in [0, 2\pi)$, in deformed theory on small  $\R^3 \times S^1 $.  
Because of  continuity, we expect all of our findings  to  hold  qualitatively for  pure Yang-Mills theory  on $\R^4$.   Arbitrary $\theta$ is problematic in lattice simulations  due to sign problem, and $N=2$ is not easy to reach using gauge/gravity correspondence.  Even if these two obstacles were not there (and we hope that in time they will be surmounted), our results provide unique  insights into  the nature of $\theta$-angle dependence.  

The main virtue of our formulation is that it interconnects seemingly unrelated topological phenomena in diverse  dimensions in  deep and beautiful ways.  We show  that the geometric (Berry) phase \cite{Berry} induced topological term in the action of certain spin systems  
\cite{Read:1990zza} and quantum dimer models \cite{Fradkin:1989hj}
is a discrete version  of $\theta$-angle in  4d gauge theory compactified on $\R^{3} \times S^1$.  This connection  can only be shown  by using compactification that respects center symmetry 
and continuity \cite{Unsal:2008ch, Shifman:2008ja}.\footnote{Using thermal compactification, the theory moves to a deconfined phase in small $S^1$, and  is  disconnected from the large-$S^1$ theory.  In this case, the connections we propose are invisible. This ``traditional" compactification  is probably the reason why the simple observations of this paper were not realized earlier.}  
A new   compactification  of gauge theory on   $T^3 \times \R$, reducing the theory to simple  quantum mechanics, shows that $\theta$ angle in gauge theory can also be mapped  to Aharonov-Bohm flux \cite{Poppitz}, and 
the interference induced by $\theta$ angle is the Euclidean realization of the 
Aharonov-Bohm effect \cite{Aharonov:1959fk}.  This provides a new perspective to  theta dependence and sign problem, and  will be discussed in a companion paper. 

Our result suggest that $\theta$-angle in 4d gauge theory  is the parent of many  topological terms   in lower dimensions.  The corresponding topological terms are inter-related and the  sign problems are  
physical,   as opposed to being technical problems.


%
\subsection{General structure of $\theta$-dependence}
The structure of the $\theta$-dependence for a subclass of   observables  in   Yang-Mills theory in the large-$N$ limit has been conjectured in 
Ref.~\cite{Witten:1980sp}  using standard {\it assumptions}  about the infrared dynamics. 
 Ref.~\cite{Witten:1980sp} argued, based on 
{\it i)}  large-$N$  't Hooft scaling  applied to holomorphic coupling 
$\tau= \frac{4\pi i}{g^2} + \frac{\theta}{ 2\pi}$,  and  {\it ii)}  the assumption that the vacuum energy density   ${E}(\theta)$ must be a $2\pi$-periodic  function of $\theta$, 
 that   ${E}(\theta)$  must be a multi-branched function: 
 \begin{equation}
 \label{largeN}
  {E}(\theta) = N^2 \min_k h \left( (2 \pi k + \theta)/N \right) \qquad \text {large-$N$}
\end{equation} 
for some function  $h$ which has a finite $O(N^0)$   limit as $N \rightarrow \infty$.  
The energy is an extensive observable which scales as $O(N^2)$, whereas the mass spectrum  scales as $O(N^0)$  in  large-$N$ limit, and is non-extensive. 
This simple observation has strong implications  for the $\theta$ dependence  of observables at large-$N$,  which are     not systematically explored in the literature. We first provide  a streamlined  field theoretic argument for general observables, and then comment on literature. 

   If we denote ${\cal H}(\theta)$  as   the  Hilbert space  of the  pure Yang-Mills theory at $\theta$, the spectrum of the theory  must obey  
    \begin{equation}
 \label{largeN3}
{\rm Spec} [{\cal H}(\theta)] = {\rm Spec} [ ({\cal H}(0)]  \qquad  \text {at $N$=}\infty
\end{equation}
We will refer to this property as {\it large-$N$ theta-independence}. A simple way to argue for 
$\theta$ independence is following. 
     \begin{figure}
\begin{center}
\includegraphics[angle=0, width=0.45\textwidth]{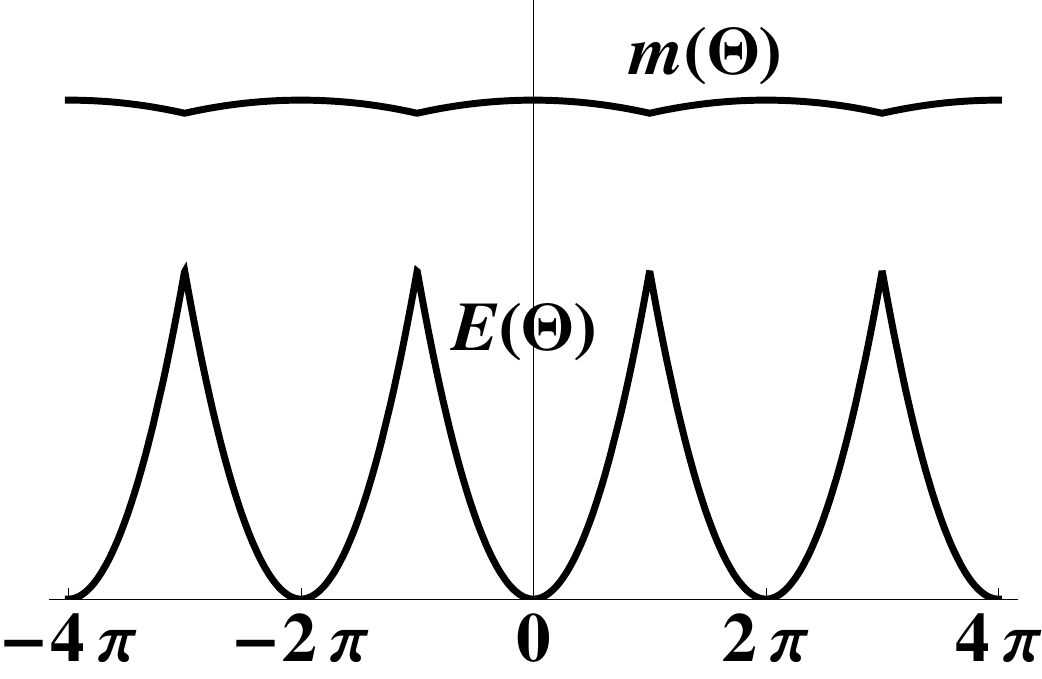}
\caption{The   $\theta$ angle (in)dependence of observables  in large-$N$ limit of gauge theory. 
For extensive observables,  such as vacuum energy density, the $\theta$ dependence is present  at $N=\infty$.  The Hilbert space and the  mass gap exhibits $\theta$ independence at 
 $N = \infty$. The figure is for  $N=5$.  At $N=\infty$, $m(\theta)$ becomes  a straight horizontal line. 
}
  \label {fig:thetadep}
\end{center}
\end{figure}

  By the assumption of a smooth large-$N$ limit,  the spectrum at $\theta=0$  is $O(N^0)$.
     Consider the mass gap associated with each branch, $m_k(\theta)$, and let $m_{k_0}(\theta)$ denote the mass gap of the theory, in the  ${\cal H}_{k_0}(\theta)$,  the Hilbert space associated with the true vacuum sector. Each branch is $2 \pi N$ periodic, but the physics is $2\pi$ periodic. 
As $\theta \rightarrow   \theta+ \psi$, for some  $\psi= O(N^0)$,  the mass of any  state in ${\cal H}_{k_0}(\theta)$ changes by an amount $O(\psi/N^{2})$. However, if $\psi=2\pi$, ${\cal H}_{k_0+1}(\theta)$ takes over as the new Hilbert space associated with the new true vacuum. Since  $O(\psi/N^{2}) \rightarrow 0$ as $N \rightarrow \infty$, the  mass gap and the spectrum of the  theory remains invariant under such shifts, implying the $\theta$-independence of non-extensive observables (\ref{largeN3}).
 Although the mass gap associated  with each branch is $\theta$-dependent, and changes drastically over the course of the full period of the particular  branch,   the spectrum of the theory built upon the true ground state, corresponding  
to the  extremum  (\ref{largeN2}),   is theta independent.

Large-$N$ $\theta$-independence  is a property of all observables which have $O(N^0)$ limits, 
 and not a property of  the extensive observables.  Specifically, the  mass gap of the theory, at  large-$N$,  ought to be 
 \begin{equation}
 \label{masstheta1}
 m(\theta)=  m(0) \max_k\left( 1 -  (\theta+ 2 \pi k)^2 \O(N^{-2})  \right), 
 \end{equation}
This implies that  the susceptibility of the mass spectrum to $\theta$-angle is $N$-dependent, it must scale as $N^{-2} $ and must vanish  at $N=\infty$. On the other hand, the topological susceptibility associated with vacuum energy density is $O(N^0)$. This leads to the 
difference in $\theta$ dependence as depicted in Fig.~\ref{fig:thetadep}.
In the opposite limit, i.e., small-$N$, 
if  (\ref{masstheta1}) approximately holds,  the mass gap and spectrum must be strongly $\theta$ dependent.    
   
 By standard large-$N$  counting, for an observable which scales as $N^{p}, p \leq 2$  in the large-$N$ limit, we expect
 \begin{equation}
 \label{largeN2}
  {\O}(\theta) = N^{p}  \;  {\rm ext}_{k}^{+} h_{\O} \left( (2 \pi k + \theta)/N \right) \qquad \text {large-$N$}
\end{equation}
for some function  $h_{\O}$ which has a finite $O(N^0)$   limit as $N \rightarrow \infty$.  
The extremum with superscript plus instructs  to choose the branch  associated with the global minimum of energy. 
 
The main message of this short description is following: The $N=\infty$ limit is 
useful to extract the theta dependence of the extensive observables. 
The same limit washes out the $\theta$ dependence of observables  which  are $O(N^0)$. 
 
 There is already  compelling lattice evidence backing  up the large-$N$   theta-(in)dependence, see for example, the structure of systematic large-$N$ expansion in Refs.\cite{Alles:1996nm,  Del Debbio:2006df, Vicari:2008jw, DelDebbio:2004ns}.   There is also evidence from gauge/gravity 
 correspondence supporting our arguments.   Ref.\cite{Witten:1998uka} shows the $\theta$ dependence of vacuum energy density in a bosonic gauge theory (which is a pure Yang-Mills theory 
 plus extra particles that  appear at the scale of glueball mass). The theta-independence of  the mass gap  is shown  in  \cite{Gabadadze:2004jq}.  
The combination of these earlier results clearly anticipates the structure of $\theta$ dependence we outlined above.

\subsection{$\theta$-dependence in  (deformed) Yang-Mills  theory}
We list the main outcomes of our semi-classical analysis  for $SU(2)$  deformed Yang-Mills theory. Because of continuity, we expect   a smooth interpolation of all physical observables 
to  pure YM on $\R^4$. 
\begin{itemize}
\item{ Mass gap, string tension and vacuum energy density are two-branched 
functions. These observables exhibit two-fold  degeneracy (and  level crossing) at  exactly $\theta=\pi$,  where they are  not smooth. The theory breaks CP spontaneously at $\theta=\pi$. } 
\item
The $\theta$ term induces a complex phase for the fugacity of topological defects.   In the Euclidean path histories and sum over  configurations in the partition function, 
these phases  generate   destructive or  constructive interference between  topological  defects. 
We refer to this phenomenon as {\it topological interference}.

\item Changing   $\theta$  radically influence  the mechanism of confinement and mass gap. 
 The mass gap at $\theta=0$ is of order $e^{-S_0/2}$ and is  due to monopole-instantons  \cite{Unsal:2008ch},  where, $S_0= \half \times \frac{ 8 \pi^2}{g^2}$ is the action of monopole-instanton, which is  half of the 4d instanton action.
At   $\theta=\pi$, the mass gap is of order $e^{-S_0}$, and   it is due to magnetic bions. 
The behavior at $\theta=\pi$ or its close vicinity is doubly-surprising, especially considering that 
 the density of monopole-instantons  $\rho_{\rm m}$ is independent of  $\theta$ angle, $\rho_{\rm m}(\theta)=\rho_{\rm m}(0)$.
 Despite the fact that    $\rho_{\rm m}$ is exponentially larger than   the  density of magnetic bions  $\rho_{\rm b}$  for  any value of $\theta$,  the  effect of the monopole instantons dies off at $\theta=\pi$ as a result of destructive topological interference. This is one of the  qualitative differences with respect to Polyakov's mechanism \cite{Polyakov:1976fu}.  
 This important effect  was missed in the earlier work by  the author and Yaffe \cite{Unsal:2008ch}.
 
 \item  The $\theta=0$ theory is  sign problem free, and $\theta \neq 0$ is a theory with a sign problem.  The corresponding sign problem is solvable by semi-classical means. The sign problem and the associated subtle cancellations may be seen as a result of topological interference. 
 \item A discrete version of $\theta$-angle  phase appears in quantum anti-ferromagnets with bipartite lattices in $d=2$ space-dimensions  \cite{Read:1990zza} and in quantum dimers \cite{Fradkin:1989hj}, as the geometric (Berry) phases.  The long distance description (a field theory on $\R^{2,1}$)  of spin-system for   $2S=0$ mod $4$ and $2S=2$ mod $4$ are equivalent,  respectively, to  $\theta=0$ and $\theta=\pi$ of deformed Yang-Mills on  $\R^3 \times S^1$. 
 The topological   $\theta$-term in YM provides a continuous generalization of the Berry phase induced term   in the spin system. 
 The existence of two vacua of the spin-system at $2S=2$ mod $4$  may be seen as 
 an evidence for CP breaking at $\theta=\pi$ in Yang-Mills. 

\item The previous  connection may seem quite  implausible  on topological grounds. The  Berry phase induced term in the spin system is   proportional,  to the first Chern number ${\rm ch}_1(B)$ associated with magnetic flux of instanton events whereas the topological term that appear in the  Yang-Mills theory is proportional to  second Chern-number, ${\rm ch}_2(F)$, the topological charge in 4d. 
 To this end, we found  a beautiful identity. In  the background
 of center-symmetric gauge holonomy, and for the topological defects pertinent to deformed Yang-Mills theory on $\R^3 \times S^1$, we show that   \footnote{
 This relation is implicitly present in my work with Poppitz \cite{Poppitz:2008hr}  on index theorem on $\R^3 \times S^1$.  
The  importance of this relation for $\theta$ dependence and dynamics  is not discussed there.}
 \begin{eqnarray}
\label{magical0}
{\rm exp}\left[{ { i \theta \;  {\rm ch}_2(F)} }  \right]= {\rm exp} \left[ { i \xi  \; \frac{ \theta}{2} \;  {\rm ch}_1(B)} \right] 
\end{eqnarray}
where $\xi= \pm 1$ for the  two different types of  magnetic charge $+1$ monopole-instanton events,   ${\cal M}_1$  and $\overline {\cal M}_2$, in deformed YM.\footnote{The existence of the second type of monopole was understood in Refs.~\cite{Lee:1997vp,Kraan:1998sn}. The role of these monopoles in semi-classical dynamics on $\R^3 \times S^1$, and in the mass gap problem and $\theta$ dependence  was initiated in Ref.~\cite{Unsal:2008ch}. }  
The opposite phases for the two same magnetic charge instanton events underlies the topological interference and  its effects on physical observables are elucidated in  
Section.~\ref{deformedYM}
\end{itemize}


\subsection{$\theta$-angle as Aharonov-Bohm effect in quantum mechanics}
Some ingredients of our formalism, especially those related to molecular instantons, 
which we also refer to as  topological molecules,    are neither widely known, nor generally correctly understood in literature. To this end,   we  decided to study 
a class of  quantum mechanical toy models as useful analogs of  gauge theory.
 These models are simple enough  to be easily tractable, but they also have enough  structure to emulate some non-trivial features of the four-dimensional counter-part. We chose to address some of the hard  issues first in this context. 

 As a simple generalization of the particle on a circle, we discuss  an infinite class of models:
A particle on a 
circle  in the presence of  a potential with  $N$-degenerate minima and a  $\theta$-term.
 For brevity,  we refer to it as  the $T_N(\theta)$-model. 
 $T_1(\theta)$  and $T_{\infty}(0)$  are well-studied text-book examples 
  \cite{ZinnJustin:2002ru, Coleman}.   
Some aspects of the   $N\geq 2$ model are parallel to  the  $SU(N)$
 dYM theory  on $\R^3 \times S^1$. 
  \begin{itemize}
 \item {  $T_N(\theta)$-model  has fractional instanton events with fractional winding number. It also has instanton events with  integer winding number.}
  \item {The physical observables are multi-branched  ($N$-branched) functions. }
    \item  {There are topological molecules, correlated instanton-instanton or instanton-anti-instanton events, topologically distinct from instantons.}
\item{The  $\theta$ angle acquires an interpretation as Aharonov-Bohm flux. 
The  $T_N(\theta)$-model can also be described as   an $N$-site lattice Hamiltonian 
 with a magnetic flux   threading  through  the ring. The topological interference due to the $\theta$-angle in the Euclidean context is the analytic continuation of the Aharonov-Bohm effect in Minkowski space.}
 \end{itemize}
 

\section{Particle on a circle}
 Consider a particle on a circle in the presence of a periodic potential and a topological $\theta$-term.  We first briefly review the standard textbook discussion of the instantons, and the semi-classical dynamics of this theory and then move to the lesser known, yet still semi-classically calculable physics of molecular instantons. 
The Euclidean action is 
\begin{eqnarray}
S^{\rm E}[g, \theta] &=& S[g]- i \theta W \cr
&=&\int d\tau  \left[\half \dot q^2_{\rm c} + g^{-1}(1- \cos q_{\rm c} \sqrt g) \right]  - i \theta\left[ 
\frac{\sqrt g}{2 \pi} \int  d\tau  \dot q_{\rm c} \right] \qquad  
\label{ac1}\\
&=&\int d\tau \;   \frac{1}{g} \left[\half \dot q^2 + (1- \cos q ) \right]  - i \theta\left[ 
\frac{1 }{2 \pi} \int  d\tau  \dot q \right] 
\label{ac2}
\qquad 
\end{eqnarray}
$g$ is the  coupling constant, which permits a semi-classical analysis for  $g\ll1$, and 
 $\theta$ is an angular variable. $W \in \Z$ is the winding number (topological term)  which depends only on the globals aspects of the field configuration.  The first form of the action (\ref{ac1}) has  a  canonically normalized  kinetic term for the field $ q_{\rm c}$, and is more suitable for perturbative discussions.  In a semi-classical analysis, it is more natural  to write the action as  in  (\ref{ac2}).

The action  $S[g]$ given in (\ref{ac2}) without any further specification is associated with {\it infinitely} many physical systems.  In order to {\it uniquely} specify the physical system under consideration, 
we have to state the configuration space of the particle, i.e., the physical identification of the 
position.  For any fixed positive integer  $N \in \Z^{+}$, we declare 
\begin{equation}
q \equiv q + 2 \pi N,   \; N \in \Z^{+},  \qquad \text{as physically the same point.}
\label{iden}
 \end{equation}
 In this section, we study $N=1$ case, for which the potential has a unique minimum within the configuration space $S^1_q$ and  the theory has a unique ground state. In this case,  $W \in \Z$ is an integer and is valued in  in first homotopy group  $\pi_1(S^1_q)=\Z$.   
 
 The general case, that we refer to as  $T_N(\theta)$-model, will be discussed in 
 Section.~\ref{TN-model}.
 

\subsection{Brief review of  instantons and dilute gas approximation}
 We first review a few well-known results in $N=1$ theory  with arbitrary $\theta$,  $T_1(\theta)$-model  in our notation, see standard textbooks \cite{ZinnJustin:2002ru, Coleman}.  
  This theory  has a unique minimum in the configuration space, $q \in [0, 2 \pi]$, and since $q$ is periodic variable, tunneling  events  $0 \rightarrow \pm 2\pi, \pm 4 \pi, \ldots$ are permitted, and present.    These   instanton effects induce a $\theta$ dependence in the ground state energy
  \begin{equation}
 E(\theta)= \half (\omega + O(g))   -2 a e^{-S_0} {\rm cos} \theta, \qquad S_0= \frac{8}{g}, \qquad  a(g)= 
 \frac{4}{\sqrt{ \pi g}}, 
 \label{gse}
 \end{equation} 
  where $S_0$ is the instanton action, and frequency of small oscillations is $\omega=1$.

  An intimately  related model is a particle moving on  an infinite  lattice $2 \pi \Z$,  in the absence of an a topological term.  This  is $T_\infty(0)$ model in our notation. In this model, there is a   $q  \rightarrow q + 2 \pi$  translation-symmetry $T$, which commutes with Hamiltonian, $[H, T]=0$. 
  There is   no physical identification between  any two lattice  points. This means, perturbatively, that there are infinitely many degenerate vacua. Non-perturbatively, this degeneracy is lifted due to tunneling events. Then,     $E(\theta)$ arises as the dispersion curve, where $\theta=k  \mathfrak{a}$ is identified as  quasi-momenta and takes all values in the interval,  $\theta \equiv k  \mathfrak{a}  \in [-\pi, \pi)$, the    Brillouin zone. The lattice spacing is labeled by $\mathfrak{a}$.
    $E(\theta=k \mathfrak{a})$  parametrizes how the infinite degeneracy of the perturbative ground states is  lifted as a function of quasi-momentum: 
    \begin{equation}
 E(k \mathfrak{a})= \half (\omega + O(g))   -2 a e^{-S_0} {\rm cos} k  \mathfrak{a} 
 \label{gse1}
 \end{equation} 
   In the $T_1(\theta)$ model,  $\theta$  is fixed for a  given theory. However, we are free to think 
 class of theories with different theta by externally  tuning it.  The ground state energy of the $T_1(\theta)$  
 model  corresponds to one of the infinitely many points in the dispersion curve of the  $T_\infty(0)$-model, using identification  $\theta=k  \mathfrak{a}$.

Let us pause for a moment, and ask a set of fairly simple, interrelated question: 
For $\theta=\frac{\pi}{2} $ (and $\frac{3\pi}{2}$),  (\ref{gse}) tells us that the dilute instanton gas 
{\it does not} contribute to the ground state energy despite the fact that the instanton 
density is independent of $\theta$.
  Why is this so?  Should we have expected this? 
 What is so special about $\theta= \pi/2$?  Will this persist at higher orders in semi-classical expansion?\footnote{ The analogous  situation in deformed YM is sufficient to appreciate 
 the importance of these simple questions. In that context,  the mass gap  at leading order in semi-classical expansion vanishes at $\theta=\pi$!    The similar  question there is whether $SU(2)$ deformed Yang-Mills, and  by continuity the ordinary YM on $\R^4$, are gapless at $\theta=\pi$?}

Consider first $\theta=0$, and the partition function $Z(\beta) = \tr [e^{- \beta H}]$ of the theory in the $\beta \rightarrow \infty$ limit, where  $Z(\beta) \sim e^{- \beta E}$. 
  In the Euclidean path integral formulation, the ground state energy receives contributions from small perturbative fluctuations around the minimum of the potential, say $q=0$,  and  from the dilute gas of   instantons corresponding to large-fluctuations: 
\begin{eqnarray}
e^{- \beta E} &\sim& e^{- \frac{\omega }{2}(1+ O(g)) \beta} \sum_{n=0}^{\infty} 
\sum_{ \n =0}^{\infty} \frac{(\beta \I)^n}{n!}  \frac{(\beta \OI)^{\n}}{\n!}  \cr
&= & e^{- \left( \frac{\omega}{2} (1+ O(g)) - \I -\OI \right)  \beta} 
\label{sum}
\end{eqnarray}
where $\I=a e^{-S_0}$  is the instanton amplitude. 

In the presence of the $\theta$-term, the instanton amplitude (or fugacity) picks up a complex phase  for each instanton event which depends  
on the $\theta$-angle as 
\begin{equation}
\I=a e^{-S_0 + i \theta}, \qquad { \overline {\cal I}}=  a e^{-S_0 -i \theta}  ~.
\end{equation}
The phases are opposite for an instanton and an anti-instanton. 
At $\theta=\pi/2$, the sum over leading instanton events  gives 
\begin{equation}
\I + { \OI} =   (e^{i \pi/2} +  e^{-i \pi/2})=0 ~.
\end{equation}
 This means, in the partition function or in their  contribution to the ground-state energy,  $\I$ and ${ \OI}$  interfere destructively.     In contrast, for example, at   $\theta=0$,  the interference is constructive. 
  This is the topological  interference which is  the source of the $\theta$ dependent structure  of observables.  Despite its  simplicity,   it leads to  qualitatively new   effects. 
 In gauge theory, we show that   topological interference effects  even alter mechanism of confinement.

 \subsection{Molecular instantons: classification}
 \label{mi}
Within the dilute instanton approximation, the vacuum energy does not receive 
any contribution at $\theta=\pi/2$. We may ask if it receives any other  non-perturbative contribution, and   if  there are  molecular (composite or correlated) 
instanton events contributing to $E(\theta)$. 
Clearly,  we must distinguish two uncorrelated  instantons and a molecular instanton.\footnote{In literature and  textbooks,   the word ``multi-instantons" is used both for  multiple uncorrelated instanton   events as well as  correlated instanton events.   In a Euclidean space, where instantons are viewed as particles,  correlated instanton events should be viewed as molecules, and carry different topological numbers than instantons.  
The role of, say, two uncorrelated instantons vs. a molecular instanton composed of two instantons  in the dynamics of the theory are completely different. This is discussed in some detail below.  
 } 
  \begin{figure}
\begin{center}
\includegraphics[angle=-90, width=1.25\textwidth]{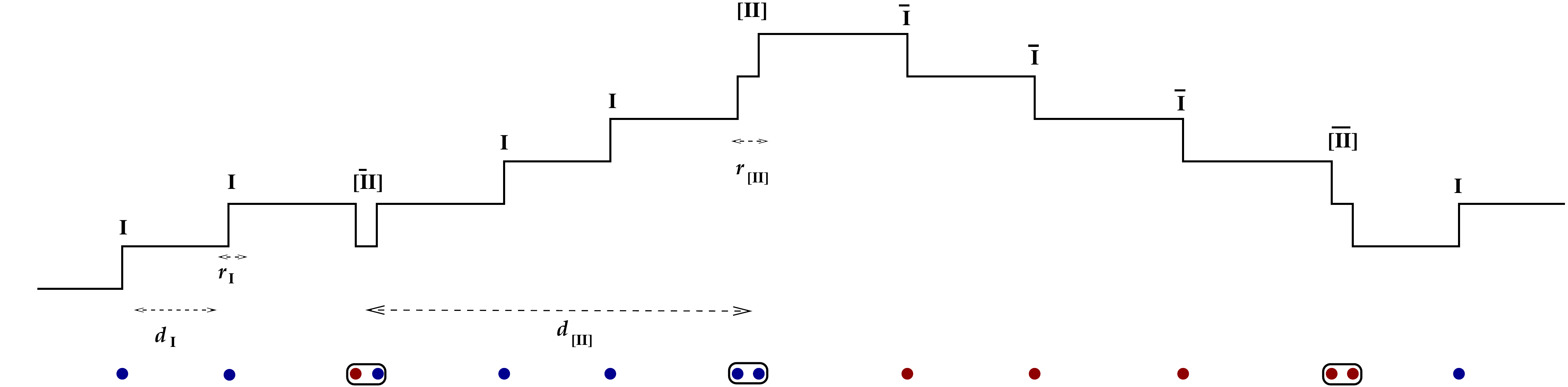}
\caption{Field configuration as a function of Euclidean time and the equivalent dilute gas of instantons and topological molecules.  
 In the textbook treatment, usually, only instantons are accounted for. Topological molecules such as $[\I \I],  [\OI \OI], [\I \OI]$ despite being rarer, are nonetheless present. There are some effects for which instantons do not contribute, and 
 the leading semi-classical contribution arise from molecular instantons. The topological molecules are  also  crucial  in  order to make sense of the continuum theory in connection with large-orders in perturbation theory. 
   }
  \label {fig:potential}
\end{center}
\end{figure}

At second order in fugacity  expansion, there are three types of molecular events: 
$[\I \OI ],  [\I\I]$, and  $[\OI\OI]$. In the Euclidean space where instanton are viewed as 
classical particles,  the correlated instanton events may be viewed as molecules. 
   We refer to molecular instanton events  with two constituents as bi-instantons, following Coleman \cite{Coleman}, and examine their properties. Much like a dilute instanton gas, we will also construct a dilute instanton, bi-instanton, {\it etc}, gas. 

The characteristic   size of the  bi-instanton molecule $r_{\rm b\I}$   is much larger  than instanton size  
$r_\I$, but  much smaller than the inter-instanton separation $d_{\rm \I -\rm \I}$ that in turn is much smaller than the inter-molecule  separation $d_{\rm b\I-\rm b\I}$. Namely, 
\begin{align}
\label{hierarchy}
\begin{matrix}
r_\I  & \ll  & r_{\rm b\I} & \ll&  d_{\rm \I-\rm \I} & \ll & d_{\rm b\I-\rm b\I} \\ 
  \downarrow   &&\downarrow&&\downarrow && \downarrow  \\
1 & \ll &  \textstyle{-  \log(\frac{g}{32})} & \ll& e^{8/g} &\ll& e^{16/g} 
\end{matrix}
\end{align}
This hierarchy  means that the use of  semi-classical  method for instantons and molecular instantons  is simultaneously  justified.\footnote{
It  is the hierarchy (\ref{hierarchy}), not the presence or absence of the  molecular/correlated  instanton events,  which is crucial  for  the validity dilute gas approximation. The presence of molecular instantons does not mean  that an instanton liquid picture needs to be used. \label{liquid} The  instanton liquid is an interesting phenomenological model,  but  obviously, it has   no  semi-classical justification. }   
We derive the size of the bi-instantons  below after we briefly  discuss their implications for the physics of the system.

The bi-instantons in $T_1$-model  are of two-types. 
\begin{itemize}
\item {\bf  $\bm {W= \pm2}$ bi-instantons:}  $ [\I\I]$ and  $[\OI\OI]$, which carry  winding number $W=\pm 2$; 
\item  {\bf  $\bm {W= \pm0}$ bi-instantons: } [$\I\OI$]  and  [$\OI\I$] which carry zero net winding number $W=0$. 
\end{itemize} 
The amplitudes associated with  $ [\I\I]$ and  $[\OI\OI]$  are given by
\begin{equation}
[\I\I]=  b(g)  e^{-2S_0  + 2 i \theta} , \qquad [\OI\OI]=  b(g)  e^{-2S_0 - 2 i \theta} 
\end{equation}
  $\pm 2$ reflects the winding number of these molecule, and $b(g)$ is a prefactor
that will be calculated in connection with  the bi-instanton  size. 
The proliferation of $ [\I\I]$ and  $[\OI\OI]$  gives a $\theta$-dependent  contribution  to $E(\theta)$, the ground state  energy. Notice that at $\theta=\pi$ where instantons interfere destructively, the bi-instanton effects are the leading non-perturbative cause of the energy shift.

 [$\I\OI$]  and  [$\OI\I$] correspond to the  amplitudes 
   \begin{equation}
[\OI\I]=  [\I\OI]   =   c(g) e^{-2S_0} \; .
\end{equation}
 $c(g)$ will  be calculated below. 
 The proliferation of these  bi-instantons   give a $\theta$-{ independent} shift to the ground state energy because  these molecules  carry zero net winding number.  
  There is in fact a  deep reason behind the 
 $\theta$ independence of $W=0$ bi-instanton contribution.  
  The perturbation theory in this simple model, despite having a  unique vacuum,  is not even 
 Borel summable, see Section.~\ref{pnp}. If one attempts to give a meaning to perturbation theory through Borel procedure, there is an  ambiguity associated with the would-be Borel sum, hence, non-summability.  The $W=0$ bi-instanton amplitude, most importantly and as will be described below, 
 is also  ambiguous, in a way to precisely cancel the ambiguity that arise from perturbation theory. 
Perturbation theory is independent of $\theta$ by  its construction and hence 
cannot mix with $W\neq 0$ sectors. By this, we mean that a contribution, say, from $W\neq 0$ sector cannot cure an ambiguity that arises in perturbation theory around the perturbative vacuum. 
 However,  perturbation theory around the perturbative vacuum can, and in fact \emph{does}, mix with non-perturbative physics in the $W=0$ sector.  
This  is the  intrinsic   difference between the two types of bi-instanton events. This will be discussed in slightly more detail in Section~\ref{pnp}, and  more fully  in a separate publication. 
 
\subsection{${W= \pm2}$ bi-instantons} 
\label{bi}
The way to derive the size of a molecule is as follows. The action of a pair of instantons is
\begin{equation}
S(z)= 2S_0 +\frac{ 32 \epsilon_1 \epsilon_2}{g} e^{-z}
\end{equation}
 where we associate $\epsilon=1$ to instantons and $\epsilon=-1$ to anti-instatons,  and $z$ is the separation between  two instanton events. The interaction is short-range  and  repulsive  for   $ \epsilon_1 \epsilon_2 =+1$ and attractive for 
  $ \epsilon_1 \epsilon_2 = -1$.  
  
   If the two instantons were {\it non-interacting}, each would  have an exact translational zero mode of its own. 
    However, instantons do interact. In this case, it is useful to split the coordinates into a relative coordinate $z= z_1 -z_2$ and center coordinate  
  $\tau= (z_1 +z_2)/2$.  The center coordinate is still an exact zero mode (as the potential between two instantons only depends on the relative coordinate) and importantly, the separation between two instantons is a quasi-zero mode, and it needs to be treated exactly.  
  
{\bf  $\bm {W= \pm2}$ bi-instantons:}   For the  $ \epsilon_1 \epsilon_2 =+1$, the integral 
$ I_{+}(g) $ over the quasi-zero mode reduce to (see   Bogomolny \cite{Bogomolny:1980ur}) 
 \begin{equation}
 \label{qzm}
 b(g)=a(g)^2  I_{+}(g),  \qquad   I_{+}(g)=\int_0^{\infty} dz \left[ e^{-\frac {32}{g} e^{-z}} -1 \right]
 \end{equation}
 The (-1) factor subtracts uncorrelated instanton events which are already taken into account in the dilute instanton approximation at the leading order. In other words, this term is there to 
 prevents the double-counting of the  uncorrelated instanton events. 
 Following Bogomolny \cite{Bogomolny:1980ur}, the interaction integral is   suppressed  in the $|z| \ll  -  \log(\frac{g}{32})$ domain due to repulsion.   
 However, the (-1) term,  which accounts for the prevention of the double-counting, 
  corresponds to the  dilute gas of instantons and  does not ``know" the repulsion.   
  Integration by parts takes care of this problem, and yields:  
\begin{equation}
\label{qzm2}
 I_{+}(g) = \frac{32}{g} \int_0^{\infty}   dz  \left[ e^{- \left( \frac {32}{g} e^{-z} +z -\log z \right)} \right]  
 = - \gamma +  \log\left(\frac{g}{32}\right) 
 \end{equation}
where   $\gamma$ is Euler constant.   Hence,  the amplitude for the bi-instanton event is 
\begin{equation}
[\I\I]= a(g)^2  \left(- \gamma +  \log\left(\frac{g}{32}\right)  \right) e^{-2S_0  + 2 i \theta} 
\end{equation}

 \begin{figure}
\begin{center}
\includegraphics[angle=00, width=0.5\textwidth]{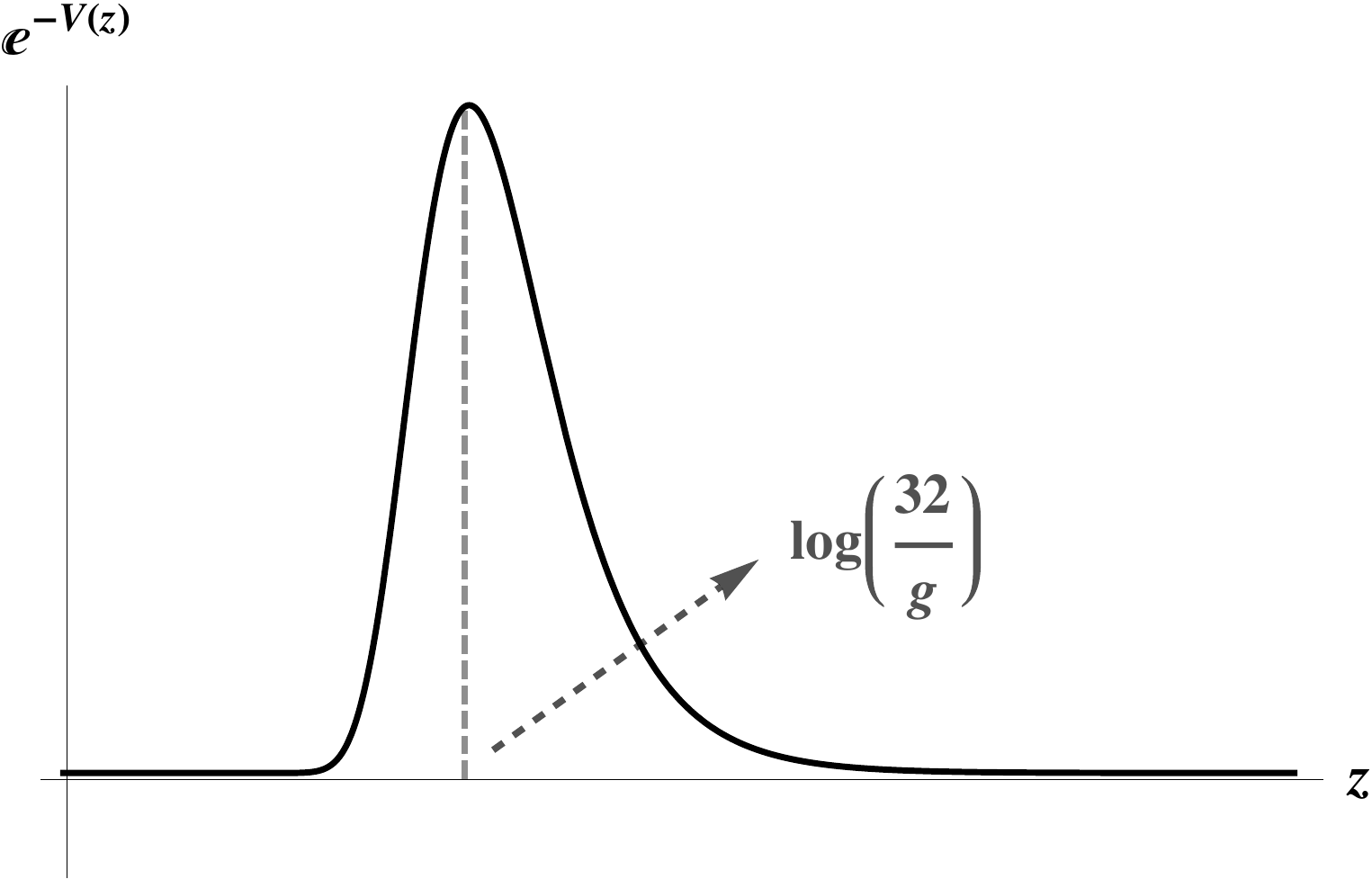}
\caption{ The plot of the integrand  over the quasi-zero mode (separation between two instanton events) for $g \ll 1$. 
 The saddle point of the integral is located at  $r_{\rm b\I}= \log\left( \frac{32}{g}\right)$.  
Since the separation between these two (correlated) instanton events  $r_{\rm b\I}$  is much larger than the instanton size, each instanton is individually  sensible. Since   $r_{\rm b\I}$ is exponentially smaller than the typical inter-instanton separation, 
these pairs cannot be viewed as  two uncorrelated single instanton events. Due to this reason, we interpret the resulting structure as a topological molecule,  with size  $r_{\rm b\I}$. }
  \label {fig:zeroQM}
\end{center}
\end{figure}

The saddle point of  the integral over the quasi-zero mode is the characteristic size of the molecular instanton event.  It is given by   $  {\it r}_{\rm b\I} \sim -  \log(\frac{g}{32})$. Clearly, the size obeys  the hierarchy  (\ref{hierarchy}).   
$ {\it r}_{\rm b}$ is much larger than  instanton size  so that each individual instanton actually makes sense,   and it is much much  smaller than inter-instanton separation so that it should be 
carefully distinguished from two uncorrelated instanton events. This characterization  is the {\it definition} of an instanton molecule.  
The existence of such molecules  do not invalidate the dilute gas approximation, 
rather  they should be accounted for.

{\bf Alternative way of evaluating  the quasi-zero mode integral:}  Another way to calculate the integral over the quasi-zero mode, which has the merit of being  straight forwardly generalizable  to quantum  field theory,  is following. Consider the theory with $f$ fermions. When $f>0$, the fermion zero-mode exchange cuts-off the  integral over the quasi-zero 
mode. This effect is familiar from the stability of magnetic bions on $\R^3 \times S^1$ 
\cite{Unsal:2007jx,arXiv:1105.0940}, and molecular instanton events in supersymmetric quantum mechanics \cite{Balitsky:1985in}.  We obtain, as the counterpart of  (\ref{qzm}), 
 \begin{equation}
 I_{+}(f, g)=\int_0^{\infty} dz e^{- \left( \frac {32}{g} e^{-z} + f z \right)}  \; .
\end{equation}
Substituting  $u=e^{-z}$ and  using $\frac{32}{g} \gg 1$,  we map this integral to 
\begin{equation}
 I_{+}(f, g)=\int_0^{1} du \;  u^{f-1} \;  e^{-   \frac {32}{g} u   }  \approx   \int_0^{\infty} du \;  u^{f-1} \;  e^{-   \frac{32}{g} u  }  = \left(\frac{g}{32}\right)^{f} \Gamma(f)
 \end{equation}
 We need $ I_{+}(\epsilon, g)$ as $\epsilon \rightarrow 0$. 
 The gamma-function  $\Gamma(f)$  has a pole at  $f=0$ zero. 
 This divergence stems from the double-counting of the uncorrelated instanton events, as described above.  Expanding the result around $\epsilon=0$, 
 we obtain 
 \begin{equation}
 I_{+}(\epsilon, g) =  \left(\frac{g}{32}\right)^{\epsilon} \Gamma(\epsilon) = \Big( 1+ \epsilon  \log  \left(\frac{g}{32}\right)   + O(\epsilon^2) \Big) \Big( \frac{1}{\epsilon} - \gamma + O(\epsilon^2)  \Big) =  \frac{1}{\epsilon}  + (\log  \left(\frac{g}{32}\right) - \gamma) + O(\epsilon)
 \end{equation}
 Our subtraction scheme,  which gets rid of double-counting of uncorrelated instanton events, is to drop the $ \frac{1}{\epsilon}$-pole term.  The result is equal to (\ref{qzm2}),   obtained earlier by Bogomolny.

\subsection{${W= 0}$ bi-instantons and Bogomolny--Zinn-Justin prescription}
 For  $ \epsilon_1 \epsilon_2 =-1$, the integral  $ I_{-}(g) $  over the quasi-zero mode is, naively, 
 \begin{equation}
 \label{naive}
 c_{\rm naive} (g)=a(g)^2  I_{-}(g),  \qquad   I_{-}(g)=\int_0^{\infty} dz \left[ e^{+\frac {32}{g} e^{-z}} -1 \right]
 \end{equation}
Now, the   interaction between the instanton and anti-instanton  is  attractive and the  integral (\ref{naive}), as it stands, is  dominated by  the  regime   $|z| \ll  -  \log(\frac{g}{32})$ where the two are close. If this is indeed the case, then   neither  the individual instanton, nor a  molecular  instanton    are   meaningful notions in the attractive case.
  In  literature, this characteristic is  sometimes regarded as unfortunate! To the contrary,  this behavior is a very positive feature, and not a bug, as described  below. Otherwise, there would  be an inconsistency in the full theory, as will be briefly described in section \ref{pnp}.

The physics of this problem is explained in two complementary papers 
by   Bogomolny and Zinn-Justin  \cite{Bogomolny:1980ur, ZinnJustin:1981dx} in quantum mechanics. Their (combined) proposal is  clever and   deep, but  as yet unappreciated in the literature. 
 Hence, we will  refer to it as  {\it Bogomolny--Zinn-Justin prescription}, or BZJ-prescription for short.  
The BZJ-prescription may  be viewed as a recipe to obtain topological molecules 
with vanishing topological numbers (just like perturbative vacuum), 
 which in the older literature are also called {\it quasi-solutions}. 
Zinn-Justin, in Ref. \cite{ ZinnJustin:2002ru}  Section 43 page 1020, states that the generalization of quasi-solutions, i.e,  topological molecules to field theory is non-trivial and has still to be worked out. 
 The present author undertook this step  in collaboration with Poppitz and Argyres.
The generalization of BZJ-prescription to non-supersymmetric quantum field theories 
on $\R^3 \times S^1$ will be given  in a detailed manner in  an upcoming work with Argyres 
~\cite{Argyres}. In Ref.\cite{arXiv:1105.3969},  it is shown that the  BZJ-prescription  produces the correct bosonic potential in a supersymmetric theory without any recourse to superpotential.

 Let us now describe the BZJ-prescription. 
 Bogomolny proposes  the following prescription in order to make sense out of attractive instanton-anti-instanton pairs. 
  Continue the coupling to negative values $g\rightarrow -g$ where the interactions between $\I$ and $\OI$ becomes repulsive, perform the integral exactly and continue back to the positive coupling. 
 The final result is $I_{+}(-g)$, or 
 \begin{eqnarray}
 \label{int2}
 c(g)&&=a(g)^2  I_{+}(-g) =  a(g)^2   \left(- \gamma +  \log\left(\frac{-g}{32}\right)\right) =   a^2  
 \left(- \gamma +  \log \left(\frac{g}{32}\right) \pm i \pi\right)   = b(g) \pm i \pi a(g)^2  \qquad \qquad 
  \end{eqnarray}
whose real part is equal to $b(g)$ given in  (\ref{qzm}).  This prescription certainly sounds  {\it ad hoc} at first.  Moreover,   (\ref{int2}) has an (naively) unexpected imaginary part proportional to two-instanton effect which is ambiguous depending on whether we approach to the positive real axis from above or below!  This results in a two-fold ambiguous $W=0$ bi-instanton amplitude:
\begin{equation}
[\I\OI]=\left( b(g) \pm i \pi a(g)^2  \right) e^{-2S_0 } 
\end{equation}
The connection of the ambiguity in the molecular amplitude   with the ambiguity that arise in large orders in  perturbation theory is explained below.

 The physical meaning of this prescription is  explained by Zinn-Justin. 
 Ref.\cite{ZinnJustin:1981dx} observes that   ordinary perturbation theory  in quantum mechanics is  divergent  for:  
 \begin{itemize}
\item[{\bf i)}]    Theories with multiple-degenerate minima. For example, 
 $V(q)= \half q^2 (1- q)^2, q \in \R$ which has two minima, and 
   $V(q)= \half (1- \cos q), q  \in \R$  which has infinitely many, or 
    $V(q)= \half (1- \cos q), q  \equiv q+ 2 \pi N, N \geq 2$, which has $N$ minima. 
\end{itemize} 
We may add  to this list
\begin{itemize}
\item[{\bf ii)}]    Theories with a {\it unique} minimum {\it and}  a periodic identifications of the fields, for example,   $V(q)= \half (1- \cos q), q \equiv q+2 \pi \in \R / 2 \pi \Z $),   
\end{itemize} 

 In this class of theories, for $g > 0$, perturbation theory is not even Borel-summable.  
 There are cases in which perturbation theory becomes Borel summable if  we take $g<0$. 
 As usual, we then {\it define} the 
 perturbative sum as the analytic continuation of the Borel sum from the negative 
 $g<0$ to $|g| \pm i \delta$. The Borel sum is well-defined  on the cut-plane, the exclusion is the  branch-cut along $g> 0$. 
 Along the branch-cut, Borel sum develops an imaginary part, which is non-unique, 
and depends on how one approaches to positive axis, from below or above,  $|g| \pm i \delta$. 
The corresponding ambiguity in the analytic continuation of Borel sum is proportional to 
$ \mp i  \pi a^2 e^{-2S_0}$. Compare this with  (\ref{int2}).

Since the ground state energy is real, the sum of perturbative and non-perturbative contributions must be real. This suggests that the imaginary part  coming from  Bogomolny prescription 
applied to winding number zero molecules  must  cancel with
the imaginary part of the Borel sum continued to the positive real axis when the two (interconnected) procedures  are performed consistently \cite{ZinnJustin:1981dx}.  Also see 
\cite{ZinnJustin:2004ib}.  

In other words,  neither the  perturbation theory on its own, nor the topologically neutral topological molecule amplitudes are unambiguous notions. Yet, the combination of the two 
must be ambiguity free. 
 
 \subsection{Validity of dilute gas approximation for  instantons and bi-instantons}
 Let  ${\cal T}=\{\I, \OI,  [\I\I],  [\OI\OI], [\I\OI],  [\OI\I],[\I\I \I],  \ldots  \} $ denote  the set of 
 instantons and molecular instantons.  The ordering is according to fugacity, the leading ones are rare and subleading ones are  rarer, but nevertheless {\it all} are present. As should be clear by now, there is also a  hierarchy (\ref{hierarchy}) of length scales. This hierarchy implies that the use of dilute 
 gas approximation which involves both instantons and bi-instantons is justified.  
 As asserted  in footnote~\ref{liquid},   the presence of molecular instantons does not mean  that an instanton liquid picture (for which there is no semi-classical justification) should  be used, much like the presence of  atoms and molecules in a gas does not imply that one should use a liquid description.

 The shift in the ground-state energy is due to the proliferation (or the grand canonical ensemble) of all defects in  set  $\T$:
   \begin{eqnarray}
e^{-E \beta} &\sim& e^{- \frac{\omega}{2} (1+ O(g)) \beta}  
\prod_{\T}  \left(\sum_{n_\T }  
 \frac{(\beta \T)^{n_{\T}}} {n_{\T}!}  \right)
  \cr 
  && = e^{- \frac{\omega}{2} (1+ O(g))  \beta}   \left(\sum_{n_\I }  
 \frac{(\beta \I)^{n_{\I}}} {n_{\I}!}  \right) \left(\sum_{n_{\OI} }  
 \frac{(\beta \OI)^{n_{\OI}}} {n_{\OI}!}  \right)  
   \left(\sum_{n_{[\I\I]} }  
 \frac{(\beta [\I \I])^{n_{[\I \I]}}} {n_{[\I \I]}!}  \right)   
 \ldots
  \cr 
&= & e^{- \left( \frac{\omega}{2} (1+ O(g))  - \I -\OI -  [\I\I] - [\OI\OI]- [\I\OI] + \ldots  \right)  \beta} 
\label{sum2}
\end{eqnarray}
 Therefore, the shift in the  ground state energy at second order in the fugacity expansion reads
\begin{equation}
 \Delta E(\theta)=  -2 a e^{-S_0} {\rm cos} \theta - 2b e^{-2S_0} {\rm cos} 2 \theta 
 - 2b  e^{-2S_0}    \;.
 \end{equation} 
At $\theta=\pi/2$, the instanton effects vanish due to destructive topological interference  and do not contribute  to ground state energy. There,  the topological molecules are  the leading  non-perturbative contribution to $\Delta E(\theta)$.

\subsection{The relation between perturbative and non-perturbative physics}
\label{pnp}
The ground state energy\footnote{This section does not aim to be complete, rather, it aims to provide the basic intuition behind  the interconnectedness  of perturbation theory and non-perturbative effects on simple physical grounds.   
The mathematical theory behind the types of series given in  (\ref{resurgent1}) and related works in mathematics and physics  literature will be covered elsewhere,  both for  quantum mechanics and quantum field theory in various dimensions, including four dimensional Yang-Mills theory.}
and eigenspectrum of the quantum mechanical system  is what is measured in an experiment and is a set of finite numbers. 
On the other hand, the perturbative expansion of ground state energy, also called Rayleigh-Schr\"odinger perturbation theory,  
 in $g$ is of the form 
 \begin{eqnarray}
 E^{(0)} (g)=  \sum_{q=0}^{\infty}   E^{(0)}_q g^q
 \label{pert}
 \end{eqnarray}
  and is  a divergent expansion, regardless of how small $g$ is. (Here, zero denotes that the calculation does not take into account any instantons or topological molecules.)  
 (\ref{pert}),   in our current example and many other cases,  is an asymptotic series. 
 By the  Poincar\'e prescription,  the series is truncated at  the minimum of the error, one obtains finite, reasonable results, with an error determined by the last term kept. However, the issue at hand is like sweeping an elephant under the rug, and  
it   does not change the fact that the series  (\ref{pert}) is actually divergent.
  Therefore,  if one takes  (\ref{pert})  literally, the perturbative expansion clashes with the finiteness of  the  ground state energy or  other observables,  meaning that, 
  a  purely  perturbative expansion to all orders is not sensible. 
 
A (still schematic) version of the  expansion for the ground state energy or  other observables 
--that may actually be given a meaning-- is  following:  
 \begin{eqnarray}
E (g) &&= E^{(0)}(g) +  E^{(1)}(g)  + E^{(2)}(g)  + E^{(3)}(g) + \ldots  \cr \cr
&&= \sum_{q=0}^{\infty}  a_{0, q} g^q +     e^{-\frac{8}{g} } \sum_{q=0}^{\infty}   a_{1,q} g^q +  
 e^{-\frac{16}{g}} \sum_{q=0}^{\infty}   a_{2,q} g^q +    
 e^{-\frac{24}{g}} \sum_{q=0 }^{\infty}   a_{3,q} g^q +   \ldots,
 \end{eqnarray} 
 where $S_0=\frac{8}{g}$ is the instanton action. 
 $E^{(1)}(g)$ is the contribution of the  dilute gas of instantons times the sum which accounts for the  perturbative fluctuations  around it, $E^{(2)}(g)$ is the contribution of the  dilute gas of bi-instantons times corresponding  perturbative fluctuations  around it, and so and so forth.  
 This expression is still  slightly incorrect, but we will correct and refine it momentarily. 

 Formally, each power series multiplying the relevant instanton factor is actually  divergent, and needs to be defined in some way. We will return to this issue in more detail later,  but in order to get a better handle on it for now, let us re-introduce the $\theta$ parameter into the expansion. This is useful because perturbation theory, by its construction,  is independent of  $\theta$-term.   
 More precisely, perturbation theory around any background, either the  perturbative vacuum or any given topological configuration, is independent of $\theta$-term. 
  This helps us to re-structure and refine the above expansion as:
  \begin{eqnarray}
E (g) &&= \sum_{q=0}^{\infty}  a_{[0,0], q} g^q   \cr 
&&+ \left[  a  e^{-\frac{8}{g}+ i \theta } \sum_{q=0}^{\infty}   a_{[1,1],q} g^q  \; + \;    a  e^{-\frac{8}{g}- i \theta } \sum_{q=0}^{\infty}   a_{[1,-1], q} g^q   \right] \cr 
&& +  \left[ a^2  \left(- \gamma +  \log\left(\frac{g}{32}\right)\right)   
 e^{-\frac{16}{g} + 2 i \theta} \sum_{q=0}^{\infty}   a_{[2,2],q} g^q     +  
   a^2  \left(- \gamma +  \log\left(-\frac{g}{32}\right)\right)  e^{-\frac{16}{g} } \sum_{q=0}^{\infty}   a_{[2,0],q} g^q \right.  \cr
&& +   \left.   a^2
 \left(- \gamma +  \log\left(\frac{g}{32}\right)\right)    e^{-\frac{16}{g} - 2 i \theta} \sum_{q=0}^{\infty}   a_{[2,-2],q} g^q  \right]
   \cr
 && +     \ldots  
 \label{resurgent1}
 \end{eqnarray} 
   The notation  $a_{[n,k],q}$  means the following: $n$ labels the action of the sector, 
 $k$ labels the $\theta$ angle dependence, or the winding number of the sector, and $q$ is a variable accounting for the perturbative expansion around a given background.  Note that the action and winding number are not necessarily proportional, and this will be  crucial in order to make sense out of such sums.  We can also define the following abbreviations for the series: 
   \begin{eqnarray}
E (g)  \equiv \sum_{n=0}^{\infty} \sum_{\substack{k=-n \\   k \rightarrow k+2 }}^{n}
E_{[n,k]} \equiv    \sum_{\substack{k=-n \\   k \rightarrow k+2 }}^{n}
  \left( {\cal Q}_{[n,k]}(g)  e^{-\frac{8n}{g}+ i k \theta }  \right)
  {\cal S}_{[n,k]},  \qquad  {\cal S}_{[n,k]}  \equiv  \sum_{q=0}^{\infty}  a_{[n,k], q} g^q \qquad
  \label{resurgent2}
 \end{eqnarray} 
 Here, $ \left( {\cal Q}_{[n,k]}(g)  e^{-\frac{n}{g}+ i k \theta }  \right) $ is the amplitude of the 
 instanton event for $n=1$ and molecular instanton event for $n \geq 2$.   ${\cal Q}_{[n,k]}(g) $ is the pre-factor of the associated instanton or molecular instanton amplitude. We have  calculated these amplitudes for $n\leq 2$. 

At least in lower dimensional theories, there is a  way  how to get a finite number out of this combined expansion, which is presumably the physical answer:  Consider the divergent  (non-Borel summable) series,  $E^{(0)} =  {\cal S}_{[0,0]}   =  \sum_{q=0}^{\infty}   a_{[0,0], q}  g^q$.  
Continue $g$ to negative $g$ in the sum. The resulting series  is Borel summable at negative $g$. Call the sum   $ 
  {\mathbb B}_{[0,0]}$.  
 $   {\mathbb B}_{[0,0]}$  is analytic function  on the cut-plane with the real positive axis excluded.  There, the function  $  {\mathbb B}_{[0,0]}  $ has an imaginary  discontinuity when passing from 
  $|g| - i \epsilon$ to     $|g| +i \epsilon$.    $ {\mathbb B}_{[0,0]}(g)= \Re  {\mathbb B}_{[0,0]} (g)]  \pm i 
  \Im  {\mathbb B}_{[0,0]} (g) $ where $   \pm i  \Im  {\mathbb B}_{[0,0]} (g)  \sim   \pm i  \pi  e^{-2 S_0}$.  
  This means that the Borel prescription for perturbation theory, as it stands, also produces 
  a  two-fold  ambiguous result, and therefore, by itself, is meaningless, because the observable 
  we are aiming to calculate is actually real.

However,  the disturbing fact that   ${\mathbb B}_{[0,0]}(g)$ produces a two-fold ambiguous result  is {\it in reality},  not in the superficial world of perturbation theory, 
 is as good as it can be.  
  Actually, without it,  we would run into an inconsistency in the whole theory. 
To see this, recall our discussion of the  proliferation of bi-instantons with $W=0$, i.e.,   the two-instanton sector associated with zero winding number, and the  BZJ-prescrription. The 
BZJ-prescription  also produces an amplitude which is two-fold ambiguous,  as in  (\ref{int2}).  Presumably, what must happen is that
\begin{equation}
 \Im  {\mathbb B}_{[0,0]} (g)   + \Im E_{[2,0]} (g)  =0 \qquad \text {up to} \; e^{-\frac{4}{g}} \; {\rm ambiguities } ,
 \label{cancel1}
\end{equation}
leading to a cancellation of the imaginary parts between the contributions coming from the $[0,0]$ sector and the contributions coming from $[2,0]$ sectors at order  $e^{-\frac{2}{g}}$. To get a finite, sensible answer for  the ground state energy,   such cancellations between the perturbative and non-perturbative 
physics must be  omni-present in the description of quantum mechanics or field theory. 
It should also be understood that the cancellation is between the $e^{-2S_0}$ effects,  the  $e^{-2S_0}$ discontinuity of the Borel  function and  $e^{-2S_0}$ imaginary part of the 
neutral bi-instanton. 
 Needless to say, there are    $e^{-4S_0}$ and lower order  imaginary contributions to the discontinuity of $  \Im  {\mathbb B}_{[0,0]} (g) $. This may potentially be cured by a molecule of the type  $[ \I \I\OI\OI]$, {\it etc.}. Hence, 
 we may expect 
 \begin{equation}
 \Im  {\mathbb B}_{[0,0]} (g)   + \Im E_{[2,0]} (g)  + \Im E_{[4,0]} (g)   + \ldots 
  =0 
 \label{cancel2}
\end{equation}
 
 We conjecture that, analogously, the same result also holds in sectors with non-zero winding number, i.e., the $\theta$ angle dependence must   also be  unambiguous: 
 \begin{equation}
 \Im  {\mathbb B}_{[1,  1]} (g)   + \Im E_{[3, 1]} (g)  + \Im E_{[5, 1]} (g)  + \ldots  =0
  \label{cancel3}
\end{equation}
In general, this suggests a recursive structure between perturbative and non-perturbative effects 
in quantum mechanics, which can be written as 
 \begin{equation}
 \Im  {\mathbb B}_{[n,  k]} (g)   + \Im E_{[n+2, k]} (g)  + \Im E_{[n+4, k]} (g)  + \ldots  =0
  \label{cancel4}
\end{equation}
Intrinsic to this cancellation is the $\theta$-independence of perturbation theory, or equivalently, the splitting of the the sectors according to winding number $k$. 
 Recall that  perturbation theory in the background of any (topological) configuration is unable to produce 
any extra   $\theta$ dependence. This means that although sectors with different action backgrounds can  mix,  the sectors with different $\theta$ dependence never mix. 
This provides a sectorial 
dynamics to the whole theory. 


  We aim to discuss the interrelation of perturbative and non-perturbative physics in quantum mechanics and quantum field theory  more systematically in the future. 
   Clearly, this is a problem of outstanding importance.

 \section{$ T_N$ -model and  fractional  winding number}
 \label{TN-model}
  For $N=1$, recall that the field $q(\tau)$ is a mapping    from the  circle  along the Euclidean time direction (with circumference $\beta$) to the target space in which the  particle lives:
 \begin{eqnarray}
 q:&& S^1_\beta \rightarrow S^1_q  \cr
  && \tau \rightarrow q(\tau)
  \end{eqnarray}
 Such  mappings are assigned  a winding number, the number of times $q(\tau)$ traverses around the   $S^1_q $ as $\tau$ makes a circuit in   $S^1_\beta$:
 \begin{equation}
 W= \frac{1}{2 \pi } \int_0^\beta  d \tau \dot q  =\frac{1}{2 \pi }( q(\beta) - q(0) ) \in \Z 
 \end{equation}
  This number depends only on  the global aspects of the field configuration, and is valued in first homotopy group  $\pi_1(S^1_q)=\Z$.  The amplitude associated with the instanton events 
  with unit winding number is   $ e^{-S_0} e^{i \theta} $.

Assume $N \geq 2$, and recall the physical identification (\ref{iden}). 
 Our assertions  about the maps   from the  circle   $S^1_\beta$   to the target space   $S^1_q $ are still  valid. The instanton interpolating from   $q(0)=0$ to    $q(\beta) = 2 \pi N$ is assigned winding number $+1$,   because $q \equiv q+ 2\pi N$  are physically the same point.
 
For convenience,  let us  normalize the circumference of the circle to $2\pi$. 
 Take the $q\equiv q+ 2\pi $ identification, but modify the potential into 
$ V(q) = 1-\cos(Nq)$. This potential has $N$-minima within the configuration space, and a $q \rightarrow q + \frac{2 \pi}{N}$ discrete shift symmetry. Let us recall the Euclidean action:
\begin{eqnarray}
S^{\rm E}[g, \theta] &=&\int d\tau \;   \frac{1}{g} \left[\half \dot q^2 + (1- \cos N q ) \right]  - i \theta\left[ 
\frac{1 }{2 \pi} \int  d\tau  \dot q \right] 
\label{acN}
\qquad 
\end{eqnarray}
We may rewrite the action in a form more suitable for instanton calculus.  
Let  ${\cal V}$ denote an auxiliary potential and   ${\cal V}' = \frac{\partial  {\cal V}}{\partial q}$ such that the bosonic potential can be expressed as $V(q) = ({\cal V}')^2$. The  auxiliary potential is the counterpart of the superpotential in supersymmetric theories.  
 Then,  the action at $\theta=0$ can be written as 
\begin{eqnarray}
S^{\rm E}[g, 0] &&=\int d\tau \;   \frac{1}{g} \left[\half \dot q^2 + \half ({\cal V}')^2
 \right]  
 = \int d\tau \;   \frac{1}{2g} \left[  \left( \dot q \pm {\cal V}' \right)^2  \mp 2 \dot q  {\cal V}'   
 \right]  \cr \cr
&&  \geq  \left| \frac{1}{g}  \int \;     d{\cal V} \right| 
 \end{eqnarray}
 where the auxiliary potential is 
 \begin{equation}
 {\cal V}= \frac{4}{N} \cos \frac{N q}{2} \; .
 \end{equation}
  The  (anti)instantons  obey
 $ \dot q \pm {\cal V}' =0$, and   saturate the bound.
 Now, there are more  possibilities for instanton  events.   
 Since there are  $N$ degenerate minima within configuration space $S^1_q$, located 
 at  $q_i = \frac{2 \pi}{N} i$,   we may view an instanton event as a tunneling event from the 
$(i)^{th}$ minimum to the $(i+1)^{th}$ minimum. Let us refer to this instanton as $\I_i$.  
The action and phase associated with this event is the integral of two total divergences, 
 $d{\cal V} $ and $d q$:
\begin{eqnarray}
S_0 -i \theta W &&=    \left| \frac{1}{g}  \int_i^{i+1} \;     d{\cal V} \; \right|  \; - \; i \theta   \int_i^{i+1} \;    d q \cr \cr
 && = \frac{4}{gN} \bigg| \cos (i+1) \pi - \cos i  \pi \bigg|  -i \theta \left( \frac{2 \pi (i+1)}{N} - \frac{2 \pi i}{N} \right) \cr \cr 
 && = \frac{8}{gN}  - i \frac{\theta}{N}
 \end{eqnarray}
This is obviously a finite action  topological configuration  whose properties depend on global aspects of the field. It  cannot be smoothly deformed to a vacuum configuration.  Such an instanton  carries    a fraction of winding number, given by $\frac{1}{N}$.
 However, this  is not valued in $\pi_1(S^1_q)$,  which is strictly  an integer. This means that  we have to relax the condition that 
 the winding number associated with an instanton event should be an integer, or refine the homotopic considerations accordingly.  
   The amplitude associated with the fractional winding  instanton is   ${\I}_i \sim  e^{-S_0} e^{i \theta/N} $.

The  discussion of  molecular instanton events follows very closely Sections~\ref{mi} and 
\ref{bi} with essentially one difference. Because of  the  ordering of the classical vacua, the interaction between instantons is modified. It is given by 
\begin{equation}
S(z)^{(i, j)} -2S_0 =  \left\{
 \begin{array}{ll} + \frac{ 32 }{g} \delta^{i, j-1} e^{-z} & \qquad \;  \text{instanton-instanton}  \\
- \frac{ 32}{g} \delta^{i, j} e^{-z} & \qquad \text{ instanton-anti-instanton}
\end{array} \right.
\end{equation}
By the same analysis as in Section~\ref{bi},  there are two-types of bi-instanton events; $W= \frac{2}{N}$ and $W=0$. These are  $[\I_i\I_{i+1}] \sim  e^{-2S_0} e^{i 2 \theta/N} $ , and $[\I_i\OI_{i}]  \sim  e^{-2S_0} $. The first one of these leads to correlated next-to-nearest neighbor tunneling, and has a $\theta$ dependence. The second one 
is an event which tunnels to the nearest-neighbor vacuum, and then immediately  tunnels back to the original vacuum. `Immediately' here means that the whole process takes a Euclidean time 
$\approx - \log \left(\frac{g}{32} \right)$, which is much larger than the instanton size, but exponentially smaller than the  separation between uncorrelated instanton events.

Note that  the winding number $W=1$ instanton event may 
be thought as an ordered concatenation  of  $N$-fractional instantons. 
The  amplitudes and the fractional winding numbers for $\I_i$ obey
\begin{equation}
  \I_{W=1} = \prod_{i=1}^{N} \I_i, \qquad  
    W= \sum_{i=1}^{N} W_i  = \sum_{i=1}^{N} \frac{1}{N} =1 
    \end{equation}
   The ${W=1}$ instanton in the $T_N$-model may be viewed as the analog of the BPST-instanton  and  the $N$ types of the  ${W=1/N}$ fractional instantons are the counterparts of the $N$-types of  monopole-instantons   on      $\R^3 \times S^1$.

We can find the $\theta$ dependence of the ground state energy   by using standard instanton methods.  Instead,  we will follow a slightly different method. 
We  map the $T_N(\theta)$-model to a $N$ -site lattice ring with a magnetic flux 
passing through the ring.

      \subsection{$\theta$-angle dependence  as  Aharonov-Bohm effect} 
      Consider the Minkowski space Lagrangian:
        \begin{eqnarray}
L[g, \theta] &=& \;   \frac{1}{g} \left[\half \dot q^2 - (1- \cos Nq ) \right]  + \frac{\theta}{2 \pi} 
  \dot q  
\label{Lag1} \qquad 
\end{eqnarray}
   The canonical momentum conjugate to the position $q$ is  $p= \frac{\partial L}{\partial (\dot q) }= \frac{\dot q}{g}+ \frac{\theta}{2 \pi} $. Thus, the Hamiltonian can be found by the Legendre transform, $H[p, q]= {\rm ext}_{\dot q} \Big[  p \dot q - L [q, \dot q ] \Big]$. 
   \begin{eqnarray}
H[g, \theta] &=& \;   \frac{g}{2}\Big( p - \frac{\theta}{2\pi}  \Big)^2  + 
\frac{1}{g}(1- \cos Nq ) 
\label{Ham1} 
\end{eqnarray} 
Therefore, the particle on a circle in the presence of the $\theta$-angle,  given in (\ref{Lag1}) and 
 (\ref{Ham1}), is same as  a charged  particle on a circle in the presence of a flux $\Phi$ treading the circle. The Aharonov-Bohm flux (in units of flux quantum $\Phi_0$) is identified with theta angle (divided by $2 \pi$):
 \begin{equation}
 \frac{\theta}{2\pi} \equiv  \frac{\Phi}{\Phi_0}, \qquad \Phi_0 \equiv \frac{2 \pi \hbar c}{|e|}
\end{equation}
This gives an experimental set-up to study the $\theta$ dependence of certain quantum mechanical systems. 

The model can possibly be studied at arbitrary coupling, $g$, however, this is not essential for our purpose.\footnote{The wave equation reduce to Mathieu or Hill's equations, for which there are known analytic solutions.} Here, our interest is the weak coupling asymptotics. 
At $g=0$, Hamiltonian reduce to the potential term. This may be viewed as an infinitely heavy particle with no dynamics,  localized at one of the minima.  At weak coupling, $g \ll 1$,  the potential term dominates, and semi-classical methods usefully apply. Below, we will solve this problem at weak coupling and study the effect of the $\theta$ term or  the magnetic flux.


  \subsection{Tight-binding Hamiltonian with Aharonov-Bohm flux}  
        \begin{figure}
\begin{center}
\includegraphics[angle=-90, width=0.7\textwidth]{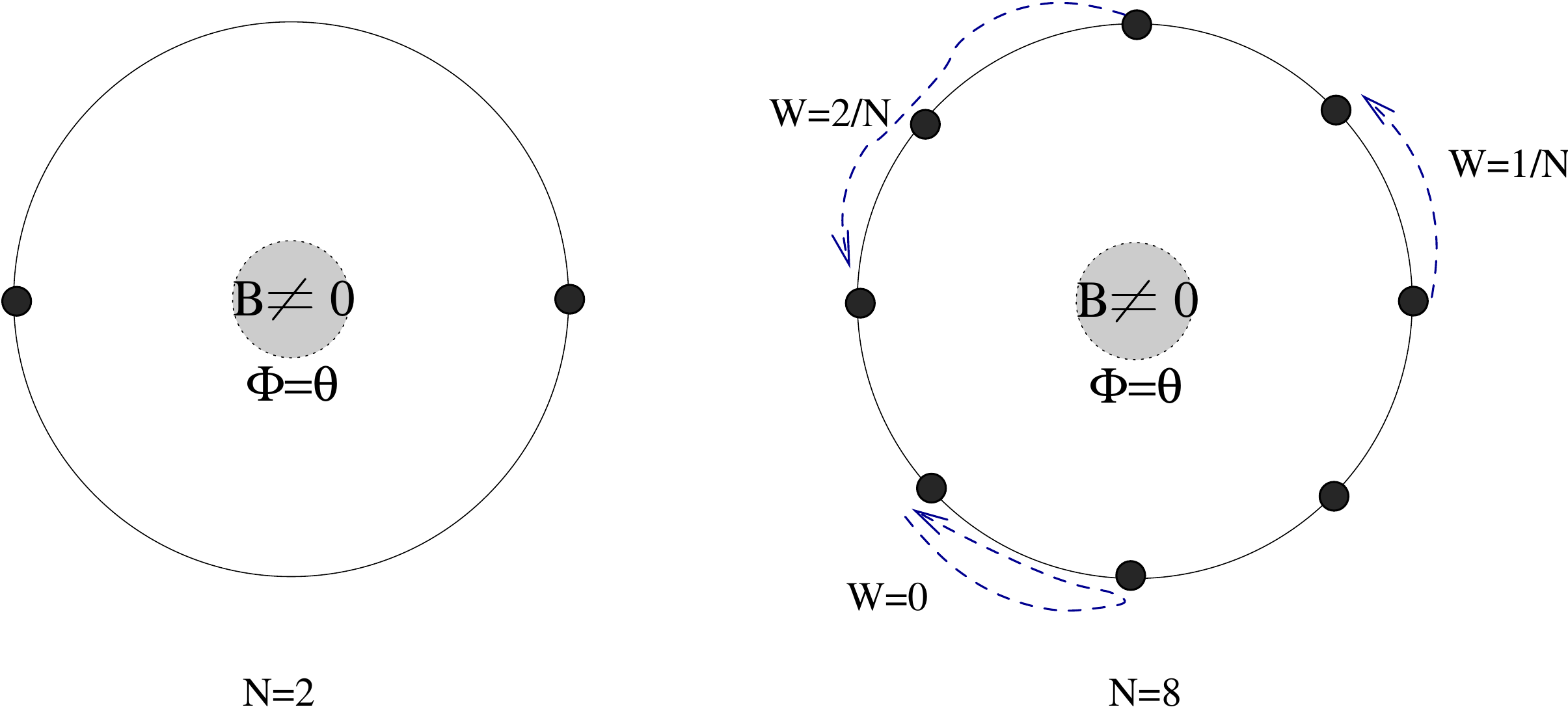}
\caption{The $\theta$ angle in the  $T_N(\theta)$-model is the equivalent of Aharonov-Bohm 
flux $\Phi$ in units of the flux quantum  $\Phi_0$, with identification  $\frac{\theta}{2\pi} \equiv \frac{\Phi}{\Phi_0}$. 
}
  \label {fig:AB}
\end{center}
\end{figure}

The $T_N(\theta)$ model  at $\theta=0$ may be approximated  by  a  one-dimensional 
tight-binding Hamiltonian $H$ on an $N$-site lattice.  The $N$ degenerate minima on  the ring  $S^1_q$ may be considered as the $N$ lattice sites. The simplest tunneling (instanton) effects correspond to nearest neighbor hoping terms in $H$.   Turning on $\theta$-angle, as described above,   is equivalent to a  magnetic flux  through the ring,  as shown in Figure \ref{fig:AB} 
 
 Let $a_j,   a_j^{\dagger}$ denote annihilation and creation operators on site $j$ obeying the canonical anti-commutation relation $[a_j, a_{j'}^{\dagger}] = \delta_{j j'}$.  
 The tight-binding Hamiltonian reads
 \begin{eqnarray}
 H =  \sum_{j=1}^{N}  \epsilon \;    a_j^{\dagger} a_j   -    t_{[1,1]}   \sum_{j=1}^{N}  \left( e^{i \theta/N}  a_{j+1}^{\dagger} a_j + e^{- i \theta/N}  a_{j-1}^{\dagger} a_j  \right)  
 \end{eqnarray}  
 where  $t_{[1,1]}e^{i \theta/N} $  is the of forward hopping amplitude and  $t_{[1,-1]}e^{-i \theta/N} $ is the backward  hopping amplitude.  The modulus of the amplitudes are equal,  
 $t_{[1,1]} = t_{[1,-1]} $, and the phase factor that particle picks up is due to the existence of Aharonov-Bohm flux. 
 
 In a Euclidean path integral formulation,   $t_{[1,1]}$ may be seen due to  simplest instanton event with positive winding number (in units of $1/N$), and    $t_{[1,-1]}$ comes from  the anti-instanton event with the same action but opposite winding. There is a directionality associated with an instanton.  
 
 The Hamiltonian commutes with discrete translation  symmetry, ${\cal T}_N$. The eigenstates obey    
   obey  ${\cal T}_N | k \rangle = e^{2 \pi i k/N}  | k \rangle$. Using the canonical transformation 
 \begin{equation}
  a_k^{\dagger}  =  \frac{1}{\sqrt N} \sum_{j=1}^{N}   e^{ 2 \pi  j k /N}   a_j^{\dagger}  \; , 
   \end{equation}  
we may diagonalize the Hamiltonian as 
 \begin{equation}
 H =  \sum_{k=1}^{N}   E_k(\theta) a_k^{\dagger} a_k  \qquad  {\rm where } \; E_k(\theta)= \epsilon  - 2 t_{[1,1]} \cos \left( \frac{\theta + 2 \pi k}{N} \right)
   \end{equation}  
$  E_k(\theta) $ is the eigen-energy of the state  $| \Psi_k \rangle $ with quasi-momentum $k $.  
 Clearly, the eigenstates   $| \Psi_k  \rangle$  are {\it independent} of $\theta$. However, the ordering of  energies depend on $\theta$.
  For the angular range $\theta \in [- \pi, \pi]$, the ground state is $k=0$, which is a translation invariant state. In the range 
 $\theta \in [\pi, 3\pi]$, the ground state is  $k=1$, which is non-singlet under the  translation  symmetry.  At $\theta=\pi$, the two states which transform differently under translation symmetry  become  degenerate and their ordering switches. This is a simple  example of a quantum phase transition where symmetry of the ground state changes as a function of an external parameter \cite{Wenbook}.  
 More generally, we have
 \begin{equation}
 \theta \in [(2k -1) \pi,  (2k +1) \pi] \; {\rm mod}(2 \pi N) \longrightarrow  | \Psi_{\rm ground} \rangle 
 = | \Psi_{k} \rangle 
  \end{equation}
  Following Ref.\cite{Gabadadze:2002ff}, we may refer to the set of states as the ``vacuum family". Every state in the vacuum family does  eventually become the true ground state 
as   $\theta$  is varied. At $\theta=(2k +1) \pi$, there is a two-fold degeneracy.     The  ground state energy (as well as the spectrum)     is a $2 \pi$ periodic function of $\theta$,  and is given by 
 \begin{equation} 
 E_{\rm g}(\theta)= \min_{k } \left[  \epsilon  - 2t_{[1,1])} \cos \left( \frac{\theta + 2 \pi k}{N} \right)  \right]
 \end{equation}
to first order in the hopping parameter expansion.  

The second order terms in the hopping parameter can be viewed as sourced by the molecular instantons.  There are two types of terms at this order, one of which has   fractional winding $\pm 2/N$ and $\theta$ dependence,   and the other  is the molecular instanton event 
with zero winding number, $W=0$ and no $\theta$ dependence. 
  We may write the second order terms in Hamiltonian as 
  \begin{eqnarray}
 H^{(2)} =   - t_{[2, 2])}   \sum_{j=1}^{N}  \left( e^{i 2\theta/N}  a_{j+2}^{\dagger} a_j + e^{- i 2\theta/N}  a_{j-2}^{\dagger} a_j  \right)    - t_{[2,0]}    \sum_{j=1}^{N}  a_{j}^{\dagger} a_j 
 \end{eqnarray}  
 Diagonalizing  the Hamiltonian, we obtain the eigen-energies of the states in the 
 vacuum family as 
 \begin{equation}
 E_k(\theta)= (\epsilon   - t_{[2,0]})    - 2 t_{[1,1]}  \cos \left( \frac{\theta + 2 \pi k}{N} \right) - 2 t_{[2,2]}  \cos 2 \left( \frac{\theta + 2 \pi k}{N} \right)
   \end{equation}  
As before, there are $N$ branches in the vacuum family, and for a given $\theta$, 
the   ground state energy is the branch with the lowest energy.


\section{Deformed Yang-Mills on $\R^3 \times S^1$ at arbitrary $\theta$} 
\label{deformedYM}
Consider Yang-Mills theory on $\R^3 \times S^1$ with action 
\begin{eqnarray}
S[g, \theta] &&= S - i \theta Q_T 
   =\int 
   \frac{1}{2g^2} \>
    \tr\,  F_{\mu \nu}^2 (x)  - i \theta 
    \frac{1}{16 \pi^2}   \int 
    \tr\,  F_{\mu \nu} \widetilde F^{\mu \nu} 
    \end{eqnarray}
   where  $F_{\mu \nu}= F_{\mu\nu}^a t^a$ is non-Abelian field strength,  $ \widetilde F^{\mu \nu}  = \half \epsilon^{\mu\nu \rho \sigma} F_{\rho \sigma}$, $g$ is 4d gauge coupling, and  $ \tr (t^a t^b ) = \half \delta^{ab}\;.$

The YM  theory possess a large-$S^1$ confined phase and small-$S^1$ deconfined phase, distinguished according to the center symmetry realization and the behavior of the Polyakov order parameter. 
There exists a  simple one-parameter family of  deformation  of the pure YM theory such that 
the deformed theory has  no phase transition as the radius is reduced. 
The  action of the   deformed Yang-Mills (dYM) theory is 
\begin{eqnarray}
S^{\rm dYM} = S -i \theta Q_T  + S_{\rm d.t.}, \qquad 
S_{\rm d.t.}= \frac{a_1}{L^4} \int  |\tr \Omega|^2. 
\end{eqnarray}
 where  $a_1$ is a judiciously chosen deformation parameter \cite{Unsal:2008ch}. 
  The small-$S^1$ regime of the deformed theory may be seen as the  {\it analytic continuation} of the confined phase to weak coupling.\footnote{The double-trace deformation  by the line operator  is only needed when $S^1$ size is smaller than the strong scale of gauge theory. 
  In this regime, this operator may be induced by introducing a heavy  adjoint fermion  endowed with periodic (not anti-periodic) boundary condition. The one-loop potential of 
  massive  fermion induce the deformation term, see \cite{Unsal:2010qh, Myers:2009df}. Since the fermion is much heavier than the strong scale, the infrared dynamics is essentially that of Yang-Mills, or equivalently, that of deformed Yang-Mills.}

  At small $S^1$, the  $SU(2)$ theory is Higgsed down to 
$U(1)$  by a center-symmetric vev $\Omega={\rm diag}\left(e^{i \pi/2}, e^{-i \pi/2}\right)$ and is amenable to semi-classical treatment. For details, see \cite{Unsal:2008ch}. Due to the ``breaking" 
$SU(2) \rightarrow U(1)$ by Wilson line, a  compact  adjoint Higgs field,  there are two types of monopole-instnantons, the regular 3d  one, and  the twisted one, which is there due to compact 
topology of adjoint Higgs, or equivalently due to the locally 4d nature of the theory 
\cite{Lee:1997vp, Kraan:1998sn}. 
These  defects carry two types of quantum numbers, magnetic and topological charge, $(Q_m, Q_T)$, given by 
\begin{eqnarray}
&&{\cal M}_{1} : (+1,  +\half), \;  \; \; \;
{\cal M}_{2} :  (-1,   + \half), \;  \cr
&&\overline {\cal M}_{1} : (-1,  - \half),  \;   \; \; \;
\overline {\cal M}_{2} : (-1,  - \half).~
\end{eqnarray}
The  action  is half of the 4d-instanton action, $S_0= \half \times S_I = \frac{4 \pi^2}{g^2}$. Note that the quantum number of ${\cal M}_{1} {\cal M}_{2}$ is the one of 4d-instanton. The $\theta=0$ theory at small-$S^1 \times \R^3$ realizes confinement due to monopole-instanton 
mechanism \cite{Unsal:2008ch}.

Introducing $\theta$ term in the action, the action of a 4d instanton is shifted as 
$S_I\rightarrow S_I - i\theta$. Since  
${\cal M}_{1}$ and ${\cal M}_{2}$ carry fractional topological charge (in a center symmetric background), and  by (\ref{magical0}), 
their action  is shifted  as  $S_0 \rightarrow S_0 - i \frac{\theta}{2}$,  whereas the shift for their conjugates is reversed, 
  $S_0 \rightarrow S_0 + i \frac{\theta}{2}$.   This is to say,  fugacities acquire  complex phases, and the  amplitudes are
\begin{eqnarray}
&&{\cal M}_{1} = a  e^{ - S_0 +  i \frac{ \theta}{2}}  e^{ +i \sigma} 
 \qquad 
{\cal M}_{2} =  a e^{ - S_0 +  i  \frac{ \theta}{2} } e^{ - i \sigma } \cr \cr
&&\overline {\cal M}_{1} = a e^{ - S_0 -  i \frac{ \theta}{2}}  e^{ -i \sigma} 
\qquad 
\overline {\cal M}_{2} = a  
 e^{ - S_0 -  i  \frac{ \theta}{2} } e^{ +i \sigma } 
  \label{2amplitudes} 
\end{eqnarray}
Here,  $\sigma$ denotes the dual photon defined through abelian duality relation,  $  \epsilon_{ \mu\nu \lambda} \partial_{\lambda} \sigma  = {4 \pi L \over g^2 } F_{\mu\nu}$. The form of the amplitudes  account for long-range Coulomb interactions between monopole-instantons. 

 \begin{figure}
\begin{center}
\includegraphics[angle=-90, width=0.9\textwidth]{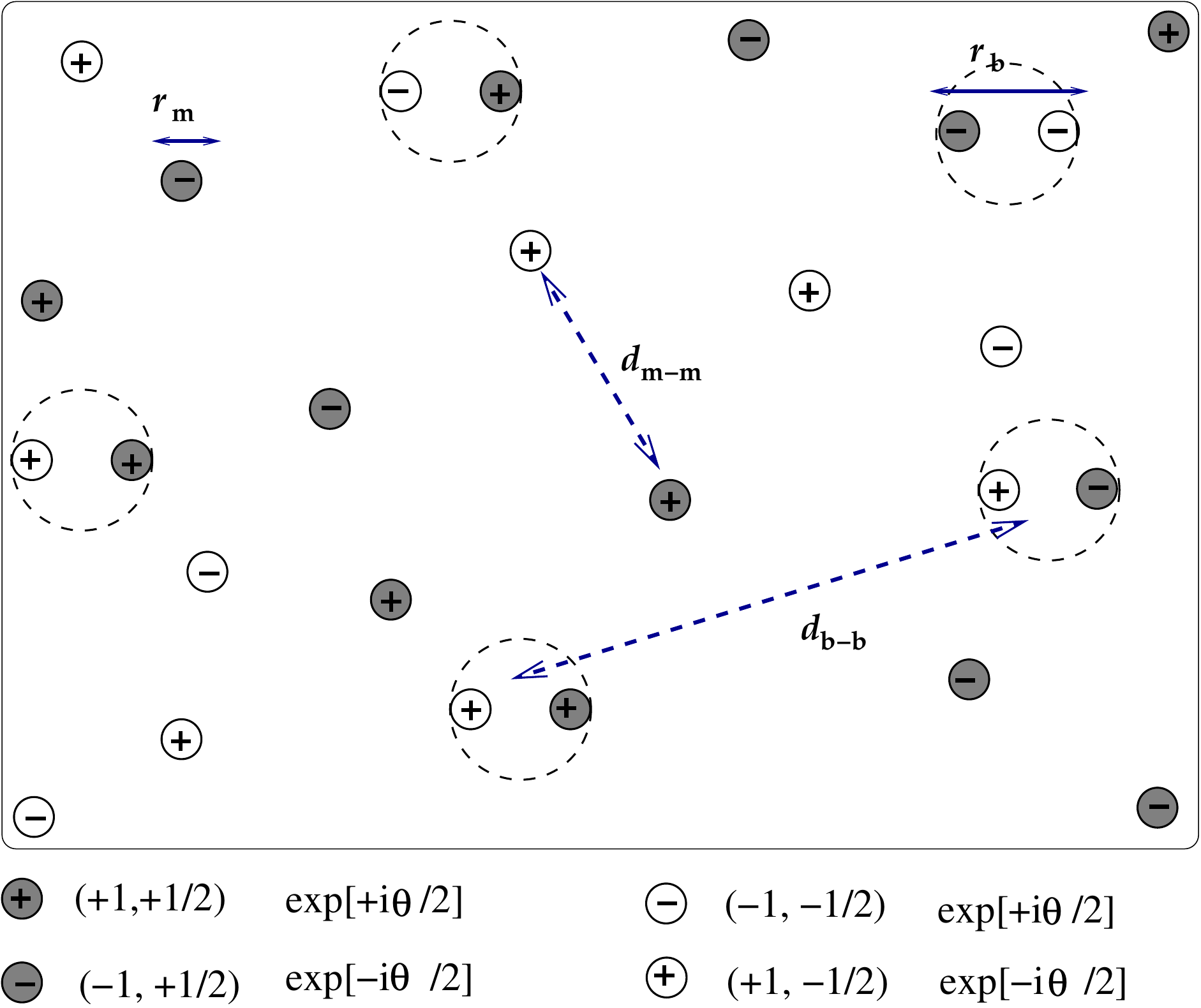}
\caption{The dilute gas of monopole-instantons and bions.  In  Euclidean space where monopole-instantons are viewed as particles, the correlated instanton events should be viewed as molecules.  Despite the fact that the density of monopole-instantons is  independent of $\theta$, 
at $\theta=\pi$,  the effect of the  monopole-instanton events dies off due to destructive topological interference, and 
the properties of dYM theory are determined by a dilute bion plasma.  }
  \label {fig:potential2}
\end{center}
\end{figure}

The dilute gas of monopoles with complex fugacity  generates the dual Lagrangian 
\begin{eqnarray}
L^{\rm d} (\sigma) =&& \frac{1}{2L} \left( \frac{g}{4\pi} \right)^2  (\nabla  \sigma)^2  - 4 a  e^{-S_0}  \cos \left( \textstyle \frac{\theta}{2} \right)   \cos \sigma  
\label{lag1}
   \end{eqnarray}
   where $V^{(1)}(\sigma, \theta)=- 4 a  e^{-S_0}  \cos \left( \textstyle \frac{\theta}{2} \right)   \cos \sigma  $  is the potential induced by the proliferation of monopole-instanton events. 
   
 For later convenience,  in order to  make the comparison to  the quantum spin system easier, 
 we  introduce a second (equivalent)  form of the dual Lagrangian, by using  the field redefinition 
 $\sigma \rightarrow \sigma  - \frac{\theta}{2} \equiv \widetilde \sigma$ . 
As a  result,  the monopole operators are modified as 
\begin{eqnarray}
&&{\cal M}_{1} = a  e^{ - S_0 }  e^{ i  \widetilde  \sigma}, \qquad 
\overline {\cal M}_{2} = a  
 e^{ - S_0 +i \theta } e^{ i \widetilde   \sigma} 
  \label{2amplitudes2} 
\end{eqnarray}
and their conjugates.  The phase differences between the two types of  monopole-instanton  events   remains the same upon field redefinition, and is a  crucial element in our discussion. The Lagrangian, in this second form, is 
\begin{eqnarray}
L^{\rm d} ( \widetilde  \sigma)  = \frac{1}{2L} \left( \frac{g}{4\pi} \right)^2  (\nabla   \widetilde  \sigma)^2  - 2 a  e^{-S_0} \Big(  \cos  \widetilde  \sigma +  \cos ( \widetilde  \sigma + \theta) \Big)  
\label{lag2}
\end{eqnarray}
 The advantage of  (\ref{lag2}) 
is its manifest  $2\pi$ periodicity. In (\ref{lag1}), to show the $2 \pi$ periodicity, one needs to use a field redefinition $\sigma' = \sigma+ \pi$ upon the shift $\theta \rightarrow \theta+ 2\pi$.

     At $\theta=0$, confinement and the mass gap for gauge fluctuations  
      are  due to   the   monopole-instantons. 
            Ref.~\cite{Unsal:2008ch}  showed that a simple generalization of Polyakov's model, which takes into account two types of monopole-instanton events, is operative in deformed Yang-Mills theory at $\theta=0$.  As we will see,  this conclusion does  not hold  for general $\theta$ due to the important topological phase (\ref{2amplitudes}). This is how the confinement mechanism presented here differs {\it qualitatively}  from Polyakov's monopole-instanton 
            mechanism \cite{Polyakov:1976fu}

     A striking phenomenon occurs at $\theta= \pi$. The monopole-instanton induced potential vanishes identically:
  \begin{equation}
V^{(1)}(\sigma, \theta=\pi)= 0 \;  
\label{vanish}
  \end{equation}
which means that  the dilute gas of monopole-instantons no longer generates a mass gap--despite the fact that their density is independent of $\theta$ angle.

 In a Euclidean volume $V_3$, 
there are, roughly,  $N_3 = V_3 \frac{  e^{-S_0}}{L^3}$ monopole events, where $L$ is the monopole size.  The monopole  density is 
$\rho_{\rm m} = N_3 /V_3 \sim   \frac{e^{-S_0}}{L^3}$, from which we can extract the mean separation between monopoles as $ d_{\rm m-m} = \rho^{-1/3}_{\rm m}  = L e^{S_0/3}$. 
Despite the fact that density of monopole does not change with $\theta$,  the mass gap at leading order in semi-classical expansion disappears. This important    effect was missed in  the earlier work of the author and Yaffe \cite{Unsal:2008ch}, and in a later work \cite{Thomas:2011ee} 
discussing the theta dependence of deformed Yang-Mills.

Experienced with the quantum mechanical example, we may guess that 
 topological interference may be taking place. This is indeed true, but there are some differences. One  may at first think   that 
${\cal M}_{i}$ must be  interfering destructively  with $ \overline {\cal M}_{i} $, for $i=1,2$. 
This is  actually not the case. 
 Since the monopole-instanton  interactions are long-ranged --- unlike instanton interactions in quantum mechanics --- the interference cannot occur between $ {\cal M}_{1}$ and 
 $ \overline {\cal M}_{1} $,  which carry opposite magnetic quantum numbers.    
 On the other hand, $ {\cal M}_{1}$ and $ \overline {\cal M}_{2} $ has the same magnetic quantum numbers, and  opposite  topological charge, see (\ref{2amplitudes}).  
At $\theta=\pi$, the sum over the   $ {\cal M}_{1}$  instanton and $ \overline {\cal M}_{2} $ anti-instanton  yields
\begin{equation}
{\cal M}_{1}|_{\theta=\pi}  + \overline {\cal M}_{2}|_{\theta=\pi}    =e^{ - S_0 } e^{ +i \sigma}  \left(  e^{i \pi/2} +  e^{-i \pi/2}\right)=0 \; ,  
\end{equation}
a destructive topological  interference, giving (\ref{vanish}). 

In order to see the  two-branched  structure of the observables in $SU(2)$ theory,  
consider (\ref{lag2}). The minima of the potential $V( \widetilde  \sigma)$ for a given $\theta$ can be found as 
\begin{equation}
\frac{d V( \widetilde  \sigma)}{d  \widetilde  \sigma}=0  \qquad \Longrightarrow   \qquad   \widetilde  \sigma =  \left\{ \begin{array}{ll}
  - \frac{\theta}{2}  & \qquad  {\rm  branch-one}\\
   - \frac{\theta}{2} + \pi  & \qquad {\rm  branch-two}
\end{array} \right.
\end{equation}
or in terms of original $\sigma =  \widetilde \sigma+ \theta/2$ field,   and potential (\ref{lag1})
\begin{equation}
\qquad \qquad  \frac{d V( \sigma)}{d   \sigma}=0  \qquad \Longrightarrow   \qquad  
\sigma =  \left\{ \begin{array}{ll}
  0 & \qquad  {\rm  minimum \; for } \;\; 0 \leq \theta < \pi  \\
  \pi & \qquad  {\rm  minimum \; for } \;\;  \pi <  \theta < 2 \pi  \\
\end{array} \right.
\end{equation}
The extremization problem has multiple solutions within the fundamental domain of $ \sigma \in [0, 2\pi)$. The nature of an extrema changes with varying $\theta$. A minimum may become a maximum or vice versa.   This results in multi-branched observables. The ground state  is associated with the branch which has lowest energy.  Various observables will be discussed in 
Section ~\ref{masstheta}.

 \subsection{Dilute gas of monopoles  and bions}
Since mass gap and  confinement   at leading order in fugacity expansion are destroyed  by topological interference, Polyakov's monopole-instanton mechanism is no longer operative. It is  natural   to ask whether confinement and mass gap will ever set in  at $\theta=\pi$, and if so, how?    

  In (deformed) Yang-Mills theory, at  $\theta=\pi$, there are  only two physical options: 
 Either the theory remains  gapless or  it has  two-fold degenerate vacua   with a 
 a much smaller mass gap, as will be shown by symmetry in  \ref{syman}.
 An identical  conundrum is recently found in  principal chiral NL$\sigma$ model  in 2+1 dimensions in Ref.~\cite{Xu:2011sj}, but was not resolved.   In gauge theory, we will be able to solve the analogous problem. 

The  question  of whether a mass gap will ever set in, or not,  is  not unfounded. 
For example, there is a well-known 
classification of  spin-$S$ antiferromagnetic spin chain in 1+1 dimensions:    half-integer spin systems are gapless, while the  integer spin systems   are gapped 
\cite{Haldane:1983ru}. 
This difference stems from a topological term in the path integral, 
$
Z (2 \pi S)= \sum_{W \in \Z } e^{i  2 \pi  S  W} Z_W
$  where $Z_W$ is the partition function over a fixed topological charge sector. Here, we may identify  $\theta \equiv 2 \pi S$ and  the crucial difference between integer spin (for which  $e^{2 \pi i S  W} =(+1)^W$) and half-integer spin (for which 
 $e^{i  2 \pi  S  W} =(-1)^W$) is the {\it signed sum} over the topological sector in the latter. 
Although this is  analogous to  the situation we encounter in dYM at $\theta=0$ vs. $\theta=\pi$, 
 we will in fact show that, despite the interference effect, a mass gap is generated. 
It is  $m^2(\theta=\pi) \sim e^{-2S_0}$,  exponentially smaller than    $m^2(\theta=0) \sim e^{-S_0}$, and the vacuum is two-fold degenerate.  This phenomenon is a generalization of what takes place in 2+1 dimensional bi-partite anti-ferromagnetic lattices  \cite{Read:1990zza}  and quantum dimer model \cite{Fradkin:1989hj}.

In order to answer the question of mass gap generation at $\theta=\pi$, 
we need to understand the topological defects at second 
order in fugacity expansion. There are  two classes of such defects, classified according to topological charge. 
These are  $[ {\cal M}_{i}  {\cal M}_{j}]$  for which  $Q_T=  1$  and $ [ {\cal M}_{i} \overline  {\cal M}_{j}]$   for which  $Q_T=0$. 
 In a normalization where the 4d instanton  amplitudes are given by 
 $ {\cal I}_{4d} = [  {\cal M}_{\rm 1}   {\cal M}_{ 2} ] =     e^{-2S_0  +i \theta }$,  and 
 $ \overline {\cal I}_{4d} = [  \overline {\cal M}_{\rm 1}    \overline {\cal M}_{ 2} ] =   e^{-2S_0  +i \theta }   $, the formal expressions for the possible 
 topological molecule amplitudes are given by  
  \begin{eqnarray}
  &&[  {\cal M}_{\rm 1}  \overline {\cal M}_{ 2} ] =b (g)  e^{-2S_0 +2i\sigma  }  \qquad  \;\;\; [{\cal M}_{\rm 2}  \overline {\cal M}_{ 1}]  =b(g)  e^{-2S_0 +2i\sigma  }    \cr  
 && [ {\cal M}_{\rm 1}  \overline {\cal M}_{ 1} ]  = c(g)e^{-2S_0}, \qquad \; \qquad 
 [  {\cal M}_{\rm 2}  \overline {\cal M}_{ 2}   ] = c (g)e^{-2S_0},  \cr  \cr
&& [  {\cal M}_{\rm 1}   {\cal M}_{ 1} ]  =d(g) e^{-2S_0+ 2 i \sigma + i \theta },   \qquad   [  \overline {\cal M}_{\rm 1}    \overline {\cal M}_{ 1} ] = d(g) e^{-2S_0- 2 i \sigma - i \theta } ,   \cr 
&& [ {\cal M}_{\rm 2}   {\cal M}_{ 2} ] =  d(g) e^{-2S_0- 2 i \sigma + i \theta },  \qquad   
[   \overline  {\cal M}_{\rm 2}    \overline {\cal M}_{ 2} ] = d(g) e^{-2S_0+ 2 i \sigma - i \theta }
  \end{eqnarray}

The molecules with $Q_T=0$ do not have a $\theta$-dependence.  $[ {\cal M}_{\rm 1}  \overline {\cal M}_{ 2}  ]$  is  capable of producing a mass gap for gauge fluctuation, as
it carries a magnetic charge plus two.  This molecule is referred to as a {\it magnetic bion} in the context of  QCD(adj)  and $\N=1$  SYM, where it is  the leading cause of confinement in semi-classical domain on $\R^3 \times S^1$ \cite{Unsal:2007jx,arXiv:1105.0940}.
  
The generalization of the analysis of  Section.~\ref{bi} can be used  to give the values of the prefactors for  the amplitudes of these  events. The result is 
\begin{eqnarray}
&&{b}(g) = \frac{2 \pi  a^2}{3}   \left(- \log\left (\frac{g^2}{4 \pi}\right) + \gamma  -\frac{11}{6} \right) 
\; , 
\end{eqnarray} 
which is the prefactor of the magnetic bion amplitude.  The analysis above is in the semi-classical domain and reliable therein. There are also lattice studies in strongly coupled domain 
providing some evidence which can possibly be interpreted in terms of  magnetic bions \cite{Bruckmann:2011cc}.

Although the $ [ {\cal M}_{\rm 1}  \overline {\cal M}_{ 1} ]$ molecule is not  important for our current  analysis,  it is of crucial importance in the full theory. In $\N=1$ SYM theory, this molecule is shown to lead to {\it center stabilization}, and is referred to as 
{\it neutral} or {\it  center-stabilizing bion}
\cite{arXiv:1105.3969}.\footnote{In order to see its role in center-symmetry,  restore the gauge  holonomy dependence in the monopole amplitude, 
$  {\cal M}_{\rm 1}  \rightarrow    e^{-\frac{4 \pi }{g^2} \Delta \phi +i\sigma  }  $, where 
$ \Delta \phi$ is the separation between two eigenvalues of Wilson line. Then,
$  [{\cal M}_{\rm 1}  \overline {\cal M}_{ 1} ] =   e^{-\frac{8 \pi }{g^2} \Delta \phi }$ leading to a repulsion between eigenvalues, and $  [{\cal M}_{\rm 2}  \overline {\cal M}_{ 2} ] =   e^{-\frac{8 \pi }{g^2} (2 \pi - \Delta \phi) }$. The sum of the two is minimized when $\Delta \phi=\pi$, the center-symmetric configuration at weak coupling regime.  See  Ref.~\cite{arXiv:1105.3969}.}
Perhaps, to keep the analogy between the molecules in quantum mechanics  and the ones in field theory as parallel as possible, we should note that the constituents of the center-stabilizing bion are also attractive. That means, we need the  generalization of the BZJ-presciption to field theory, which is undertaken in  \cite{Argyres}.   Following  Ref.~\cite{Argyres}, we find, 
\begin{eqnarray}
&&{c}(g) = \frac{2 \pi  a^2}{3}   \left(- \log\left (-\frac{g^2}{4 \pi}\right) + \gamma  -\frac{11}{6} \right)  
= b(g) \pm  \frac{2 \pi  a^2}{3} ( i \pi)
\end{eqnarray} 
As in quantum mechanics, the (refined) BZJ-prescription leads to  an   imaginary part contribution to vacuum energy. In Yang-Mills  theory, we also expect that the vacuum energy in perturbation theory to be  non-Borel summable.  In order for the gauge theory to make sense, the ambiguity (associated with non-Borel summability) must cancel with the two-fold ambiguity of the 
 neutral  bion contribution.\footnote{ This molecule is associated with a pole in the Borel plane 
 at $t=8 \pi^2  = \half (16\pi^2)$, where $t=16\pi^2 $ is the pole corresponding  to 4d instanton-anti-instanton. 
Ref.~\cite{Argyres} provides evidence that  the  neutral bion is the weak coupling semi-classical incarnation  of the elusive IR-renormalon (for which, up to our knowledge, no semi-classical description exists.)   We are  quickly glossing over this issue here, for the fuller discussion, see \cite{Argyres}}

   The characteristic   size of the  
 $   [ {\cal M}_{i} \overline  {\cal M}_{j}] $ molecules can be found, as in quantum mechanics, 
 by studying the   integral over the quasi-zero mode.   The result is, parametrically,   
 $r_{\rm b} \sim \frac{L}{g^2}$, same as the magnetic bion size in QCD(adj) or  $\N=1$ SYM 
 \cite{Unsal:2007jx, arXiv:1105.0940},  and is  universal. 
The bion size  is much larger  than monopole-instanton size  $r_{\rm m} \sim L$, but  much smaller than the inter-monopole separation $d_{\rm m -m}\sim L e^{S_0/3}$ that in turn is much smaller than the inter-bion separation $d_{\rm b-b}  \sim  L e^{2S_0/3}  $.  
Namely, 
\begin{align}
\label{hierarchy2}
\begin{matrix}
r_{\rm m}  & \ll  & r_{\rm b} & \ll&  d_{\rm m-m} & \ll & d_{\rm b-b} \\ 
  \downarrow   &&\downarrow&&\downarrow && \downarrow  \\
  L &  \ll & \frac{L}{g^2}  & \ll & Le^{S_0/3} & \ll &  Le^{2S_0/3}
\end{matrix}
\end{align}
Again,  this hierarchy  means that the use of  semi-classical  methods for a dilute gas of 
instantons, bions, and other topological molecules is simultaneously justified.

On the other hand, the molecules appearing in the first class have non-universal properties. 
Whether these molecules form or not depends on the details of theory. In dYM, their properties are dependent on the mass of $A_4$-scalar, and hence on the deformation parameter $a_1$.   The characteristic  $A_4$-mass in the center-symmetric phase  is $\frac{g}{L}(a_1-1)$. If $m_{A_4}=0$,  the net interaction between  self-dual  monopole-instantons vanishes:  the $\sigma$-scalar  exchange is  cancelled by the  $A_4$-scalar  exchange.  This is unlike bions, where the interaction strength is  parametrically unaltered  in the limit $m_{A_4}=0$.  The size of the bion is 
only altered   by a factor of two in this limit.  For a range of  $a_1$ 
deformation  parameter, the amplitude associated with  the $Q_T=1$   type  events are  much suppressed $d (g) \ll b(g) $ relative to $Q_T=0$ events. This approximation becomes exact in the supersymmetric $\N=1$ theory, as well as its softly broken 
 $\N=0$ non-supersymmetric version. This suggests that we can omit such events with respect to bions in the long-distance description and we will do so.


Let  ${\cal T}=\{ {\cal M}_{i},  \overline {\cal M}_{i} ,  [ {\cal M}_{i}  
 \overline {\cal M}_{j}],   [ {\cal M}_{i}  {\cal M}_{j}]  \ldots  \}
  $ denote the set of  topological defects and molecules in dYM.  The grand canonical partition function of this Coulomb gas is
\begin{equation}
    Z =
    \prod_{\T} 
    \left\{
    \sum_{n_{\T}=0}^\infty
    \frac {(\zeta_\T)^{n_{\T}}}{n_{\T}!}
    \int_{\R^3}
    \prod_{k=1}^{n_{\T}} d{\bf r}^{\T}_k  \right\} 
     e^{-S_{\rm int} ({\bf r}^{\T}_k ) } \,,
\label{eq:grand}
\end{equation}
where $S_{\rm int}$ denotes the   Coulomb interactions among the set of defects in   ${\cal T}$, 
and $\zeta_\T$ is the fugacity of ${\cal T}$. 
Unlike  Ref.\cite{Unsal:2008ch}, which only took into account the monopole-instantons in the  compactified theory, we   also  include  the defects at  second order in the semi-classical expansion.  This is necessary (and sufficient)  to correctly describe the  infrared physics at arbitrary $\theta$ in the small $S^1 \times \R^3$ domain.   We do keep 
the BPST instanton induced term in the action,  not because it should be kept to capture the long-distance physics correctly,  rather to show its 
unimportance of its contribution to observables. 
The  partition function  can be  transformed into a $3d$ scalar
field theory 
$
    Z (\theta)  =
    \int  \mathcal D\sigma \;
    e^{-  \int_{\R^3} 
    L_{\rm d}[\sigma]} 
    $ 
where 
\begin{eqnarray}	
L^{\rm d} = \frac{1}{2L}
	\left(\frac{g}{4\pi}\right)^2
	(\nabla \sigma)^2  -  
 \underbrace{4 a e^{-S_0} \cos \textstyle \left(\frac{\theta}{2} \right)   \cos \sigma }_{\rm monopole-instanton}  -  
  \underbrace{2b  e^{-2S_0}    \cos 2  \sigma}_{\rm magnetic \; bion}   -  \underbrace{2 a_{\rm 4d} e^{-2S_0}  \cos \theta}_{\rm BPST-instanton}
\label{pot}
  \end{eqnarray}
 The physical  aspects of the long-distance  theory are  captured by this dual action  
(\ref{pot}).  These are examined below. 

In order to make the correspondence with quantum anti-ferromagnet easier,  we will also give the equivalent  Lagrangian  in terms of shifted variable $ \widetilde \sigma =  \sigma  - \frac{\theta}{2}$. It is 
\begin{eqnarray}
L^{\rm d} ( \widetilde  \sigma)  = \frac{1}{2L} \left( \frac{g}{4\pi} \right)^2  (\nabla   \widetilde  \sigma)^2  - 2 a  e^{-S_0} \Big(  \cos  \widetilde  \sigma +  \cos ( \widetilde  \sigma + \theta) \Big)   
- 2b  e^{-2S_0}    \cos( 2 \widetilde \sigma + \theta) - 2 a_{\rm 4d} e^{-2S_0}  \cos \theta \qquad \qquad
\label{lag2mod}
\end{eqnarray}

\subsection{Vacuum energy density and topological susceptibility}
\label{masstheta}

\begin{figure}[ht]
\begin{center}
\includegraphics[angle=0, width=0.60\textwidth]{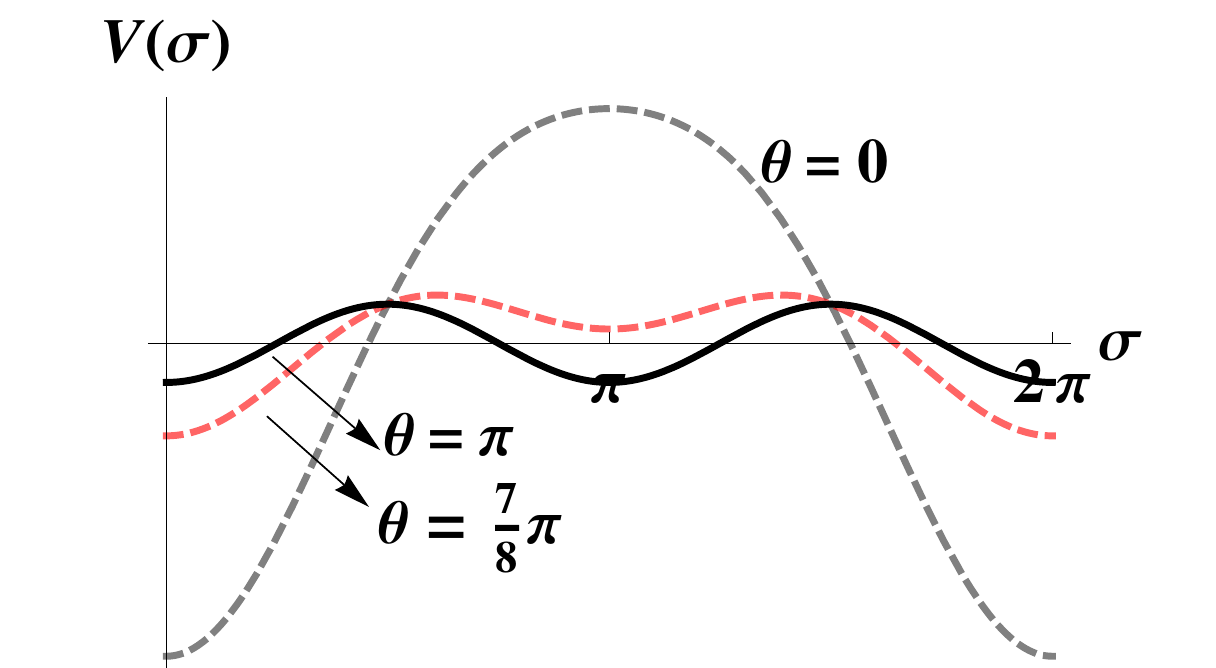}
\caption{$V(\sigma, \theta)$ as a function of $\sigma$ for $\theta=0, \frac{7\pi}{8}, \pi$. At $\theta=0$, there is  a unique ground state. For a range of $\theta$, there are two minima. 
 At $\theta=\pi$, there are   two degenerate  (ground) states. 
 }
  \label {fig:potential3}
\end{center}
\end{figure}

The potential (\ref{pot}),  for arbitrary $\theta$, has {\it two} $\theta$-{\it independent} 
extrema,  located at  $\sigma= \{0,  \pi\}$, which lead to two competing vacua. 
There are also, for  a range of $\theta$, two $\theta$-dependent extrema.  But these are  always maxima. 
The ``vacuum family", in the sense of Ref.\cite{Gabadadze:2002ff} is captured  by theta-independent extrema  of (\ref{pot}),  at least one of  which  is always 
a minima.  For a range of $\theta$, there are two minima,  located at  $\sigma= 0, {\rm and} \;  \pi$, independent of $\theta$. 
See the potential for dual photon, Fig.~\ref{fig:potential3}, for three values of $\theta$.   

 Because  of the existence of  two candidate vacuum states, 
physical observables, such as vacuum energy density, mass gap, string tension, deconfinement temperature  are two-branched functions.  Because the  two   candidate ground states become degenerate  at $\theta=\pi$, or at odd-multiples of $\pi$, the observables are  smooth except 
for  odd-multiples of $\pi$, where it is non-analytic.

The true ground state properties, for a given  $\theta$, are  found by using the branch associated with the global minimum of energy. 
The vacuum energy density ${\cal E} (\theta)$ is extracted from the value of the 
$V(\sigma, \theta)$ evaluated at these two extrema, 
$L {\cal E} (\theta) = {\rm Min}_{k=0,1} \left[ V(k \pi, \theta) \right] $. Explicitly, 
\begin{eqnarray}
{\cal E} (\theta)  =  \Lambda^4 \; \min_{k=0,1} \left[ -4 a  (
\Lambda L)^{-1/3} \cos \left( \frac{\theta + 2 \pi k }{2}  \right)  -2 b  (\Lambda L)^{10/3}  -2 a_{\rm 4d}  (\Lambda L)^{10/3}   \cos \theta  + \ldots \right] \qquad 
\end{eqnarray}

\begin{figure}[ht]
\begin{center}
\hspace{0.5cm}\includegraphics[width=15em]{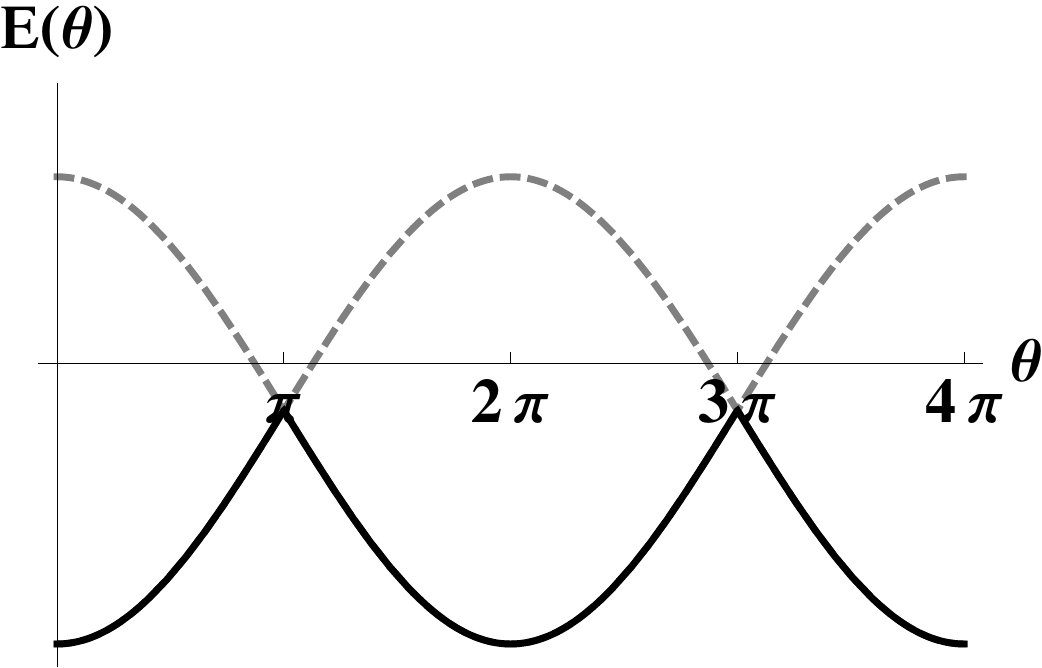} \qquad \qquad \qquad
\includegraphics[width=15em]{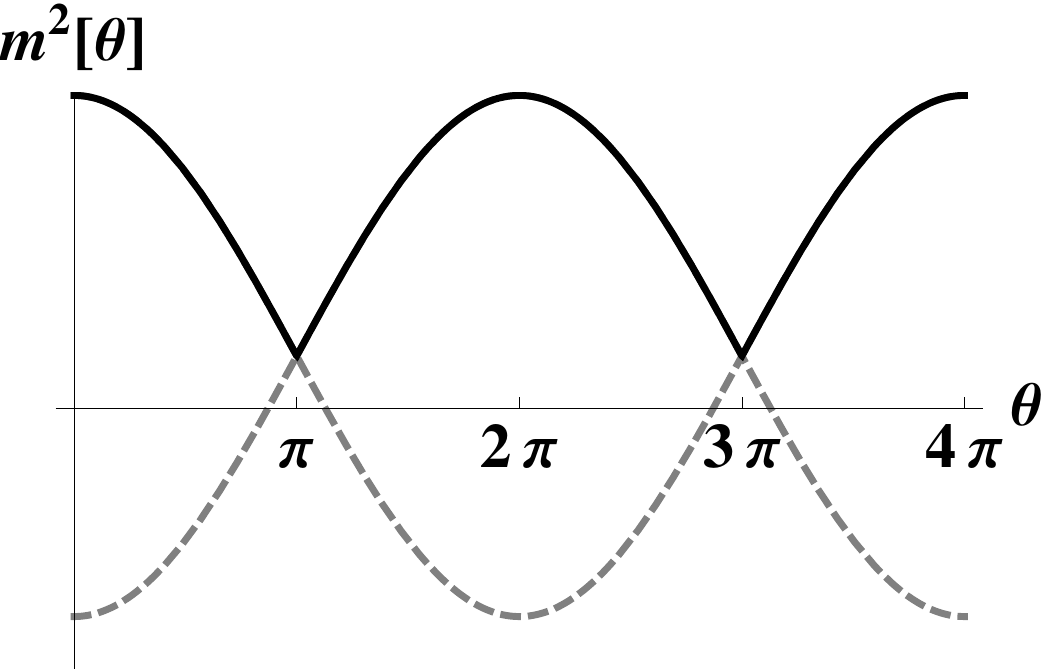}
\caption{a) The vacuum energy density $E(\theta)$ is periodic by $2 \pi$ and smooth except  for odd-
multiples of $\theta=\pi$, where a  two-fold degeneracy arises. 
b)The mass gap  of the theory, associated with the global minimum of vacuum energy,   is  the   maximum  of the two branches. At $\theta=\pi$, there is spectral degeneracy. 
 \label{fig-theta}}
\end{center}
\end{figure}

Recall that the multi-branch structure is a  conjecture on   $\R^4$  for large-$N$ theory
\cite{Witten:1980sp}. Here, we were able to derive the  two-branched structure, shown in Fig.~\ref{fig-theta} starting with microscopic  physics in a  semi-classical framework 
in deformed Yang-Mills theory.  By continuity, we expect that this result also holds for
pure  Yang-Mills theory on $\R^4$. 

 The multi-branched structure is sourced by topological defects with fractional topological charge.   It is also worth noting that the  4d-BPST instanton effects in this expansion are analytic, negligible and unimportant. 

 We can also extract topological susceptibility:
  \begin{equation}
  \chi= \frac{\partial^2 {\cal E}}{ \partial \theta^2} \Big |_{\theta=0}   \approx  
   \Lambda^4  a  (\Lambda L)^{-1/3}    + 2 a_{\rm 4d}  (\Lambda L)^{10/3}  + \ldots  
   \end{equation}
  The crucial point in this expression is that  the 4d BPST instanton effects, even in the semi-classical domain, give negligible contributions to topological susceptibility. 
  This  is   in accordance with lattice results \cite{Alles:1996nm,Vicari:2008jw}. In the semi-classical regime, in (deformed) YM theory, the leading contributions are from monopole-instanton events.

\subsection{Mass gap, string tension and deconfinement temperature}
The mass gap of the theory is also a two-branched function. It can be extracted from the 
curvature of the potential at its minima:
 $m_{1,2}^2 (\theta) = 
  L \left(\frac{ 4 \pi}{g}\right)^2 \frac{\partial^2  V (\sigma, \theta) }{ \partial \sigma^2}|_{\sigma=0, \pi} $.  
At leading order in the semi-classical expansion, we find   
      \begin{equation} 
      m(\theta) = A  \Lambda (\Lambda L)^{5/6} \Big | \cos \textstyle \left(\frac{\theta}{2} \right) \Big |^{1/2} 
      \end{equation}
 At leading order in semi-classical expansion,  at $\theta=\pi$, mass
 gap vanishes despite the fact that the density of monopole-instantons is independent 
 of $\theta$. This is a consequence of destructive topological interference.     At this stage, the theory has two choices, either to  remain gapless or two have to isolated gapped vacua. A similar problem also appears in Refs.\cite{Xu:2011sj, Senthil}.  
  At subleading $e^{-2S_0}$ order, a much smaller mass gap is  generated due to magnetic bions, and it is proportional to
$   m(\pi) \sim   \Lambda (\Lambda L)^{8/3}$. 

The mass gap of the theory is   the upper branch of a two-branched function:
\begin{equation} 
 m^2 (\theta) =   \max_{k=0,1} a \Lambda^2 \left[   (\Lambda L)^{5/3} \cos   \left( \frac{\theta + 2 \pi k }{2}  \right)  +     (\Lambda L)^{16/3}  + \ldots \right]
\label{mgth}
 \end{equation}
   For the range of $\theta$ for which  both  $m_1^2>0$ and $m_2^2>0$, there are two minima. If  $m_1^2>0$ and $m_2^2<0$ (or vice versa),  then the second extremum is actually a maximum. 
The functions ${\cal E} (\theta)$ and  the mass gap  are smooth function for all $\theta$, but odd multiples of $\pi$, where they are non-analytic.
At these values,   there are two true ground states, located at $\sigma=0$ and $\sigma=\pi$. This is a  manifestation of the CP-symmetry at $\theta=\pi$, which is spontaneously broken, and is discussed in \ref{syman}.

We also define the  topological susceptibility of the mass gap (square) as  
\begin{equation} 
\chi_m = \frac{\partial^2 [m^2(\theta)]}{\partial \theta^2} \Big |_{\theta=0}= -  A \Lambda^2    (\Lambda L)^{5/6}/4 <0
 \end{equation}
This implies that at  $\theta=0$, the  mass gap is maximum (the correlation length is minimum). With increasing theta, due to the topological interference of the monopole-instantons,  the mass gap decreases and  correlation length increases.  Although we have not been able to do so yet, we believe that it can be proven rigorously that the mass gap (and spectrum) susceptibility is negative semi-definite: 
It is negative for all finite $N$ for $SU(N)$ and approaches zero at $N=\infty$ limit. It may be interesting to demonstrate this analytically and check it by using lattice techniques. For example, 
a recent lattice work  \cite{Bogli:2011aa} studies mass gap in two-dimensional $O(3)$  field theory at arbitrary $\theta$ and claims that this should be feasible for $SU(2)$ Yang-Mills theory. It would be interesting to check (\ref{mgth})  through simulations.

{\it String tension:} The string tension may be evaluated by calculating the  expectation values of large Wilson loops in the defining $\half$-representation of $SU(2)$,   $\big\langle W_{1/2} (C)\big\rangle$. This calculation is done for deformed Yang-Mills theory at $\theta=0$  in \cite{Unsal:2008ch}. We refer the reader there for details, and here we mainly quote the differences. 
$\big\langle W_{1/2} (C)\big\rangle$  is expected to decrease exponentially with the area of the minimal spanning surface,
\begin{equation}
    \big\langle W_{1/2} (C)\big\rangle
    \sim
    e^{-T_{1/2} (\theta) \> {\rm Area}(\Sigma) } \,.
\end{equation}
Here $\Sigma$ denotes the minimal surface with boundary $C$,
and $T_{1/2} (\theta) $ is the  $\theta$-dependent  string tension for $\half$-representation.
Such area law behavior implies the presence of an asymptotically
linear confining potential
between static charges  in  $\half$-representation, 
$V_{\mathcal R}(\x) \sim T_{1/2}(\theta) \, |\x|$ as $|\x| \to \infty$.

The insertion of a Wilson loop $W_{1/2}(C)$ in the original theory 
corresponds, in the low-energy dual theory,
to the requirement that the dual scalar fields have non-trivial monodromy,
\begin{equation}
    \int_{C'} d \sigma = 4 \pi \times (\half) =  2\pi\,,
\label{eq:monodromy}
\end{equation}
where $C'$ is any closed curve whose linking number with $C$ is one. For an $\R^2$ filling Wilson loop in the $xy$-plane,  this is equivalent to finding the action of the kink solution interpolating between $\sigma=0$ at $z=-\infty$ and  $\sigma=2\pi$ at $z=+\infty$. 
At leading order in semi-classical expansion, we find, 
\begin{equation}
T_{1/2}(\theta) \sim \Lambda^2 (\Lambda L)^{-1/6} \Big | \cos \textstyle \left(\frac{\theta}{2} \right) \Big |^{1/2} 
\end{equation}
 Clearly,  $T(\theta + 2 \pi) = T(\theta)$. At $\theta=\pi$, the string tension vanishes at leading order in semi-classical expansion just like the mass gap did. 
 This means that at $\theta=\pi$, and at leading order in semi-classical expansion, the gauge theory does not confine.  
 However, at subleading ($e^{-2S_0}$) order, a much smaller string tension is generated due to magnetic bions.  The string tension at $\theta=\pi$ is, 
 \begin{equation}
   T(\pi) \sim   \Lambda^2 (\Lambda L)^{5/3}
\end{equation} 

We may also discuss the susceptibility of the string tension to the $\theta$ angle, $\chi_T = \frac{\partial^2 T(\theta)}{\partial \theta^2} \Big |_{\theta=0}$. 
 The conclusions are quite similar to the ones for the mass gap.  Most importantly, the susceptibility is negative 
 for $SU(2)$.  
Since the string tension is a non-extensive observable, the susceptibility must 
reach  zero as $N \rightarrow \infty$. In other words, the string tension at $N=\infty$ must be  
$\theta$-independent, as per our discussion in Section~\ref{sec:level1}.

{\it Deconfinement temperature:}
Consider the deformed YM on $\R^3 \times S^1_L$, where we inserted the subscript $L$ to remind the reader that there is a deformation along this circle, and the theory at any value of $L$ 
is confining.  In the small-$L$ regime, we can examine the deconfinement transition by semi-classical techniques by  introducing  a thermal thermal circle $S^1_\beta$ (with no deformation), and considering the theory on  $\R^2 \times S^1_L \times S^1_\beta$.  
At $\theta=0$,   the physics near the deconfinement temperature    is described by a classical 2d XY-spin model with a $U(1) \rightarrow \Z_2$-breaking perturbation, and 
the transition temperature is, in the semi-classical domain, 
$\beta_{\rm d} (\theta=0) = \frac{4 \pi L}{g^2}$ \cite{Anber:2011gn}. 
At $\theta=\pi$,  according to ($\ref{pot}$), the monopole effects disappear. If we do not incorporate the  magnetic bion term, the theory does not confine, i.e., the theory is then in the deconfined phase for any $T \geq 0$. Incorporating magnetic bions,  for sufficiently low temperatures the theory is confined, but we expect the deconfinement temperature to be reduced with respect to $\theta=0$ case.  At $\theta=\pi$, 
    the physics near the deconfinement temperature    is described by a classical 2d XY-spin model with a $U(1) \rightarrow \Z_4$-breaking perturbation. This is same as  $SU(2)$ QCD(adj) discussed in 
\cite{Anber:2011gn}. In this latter case,  $\beta_{\rm d} (\theta=\pi) = \frac{8 \pi L}{g^2}  =2\beta_{\rm d} (\theta=0)  $. Therefore, in terms of temperatures, 
\begin{equation}
T_{\rm d}(\theta=\pi) = \frac{1}{2}  \; T_{\rm d}(\theta=0) 
\end{equation}
To calculate $T_{\rm d}(\theta)$ for general $\theta$  is a more demanding task, but it is possible by using the RG techniques described in \cite{Anber:2011gn}. As mentioned above, on physical grounds, we should expect a lower deconfinement temperature at $\theta=\pi$ and indeed, this is 
the case. 

Finally, in the large-$N$ limit, the deconfinement temperature must exhibit $\theta$ independence
because it is a non-extensive observable,  as per our discussion in Section~\ref{sec:level1}.

\subsection{CP-symmetry and its realization} 
\label{syman}
In the microscopic theory, under CP,  $e^{- i \theta    \frac{1}{16 \pi^2}   \int     \tr\,  F_{\mu \nu} \widetilde F^{\mu \nu} } \rightarrow e^{+ i \theta     \frac{1}{16 \pi^2}   \int     \tr\,  F_{\mu \nu} \widetilde F^{\mu \nu} } $. Since $\theta$ is $2\pi$ periodic and the second Chern number is an integer for 4d instanton configurations,   CP is a (non-trivial) symmetry of the theory  if and only if  $\theta=\pi$,  because $-\pi  + 2 \pi =\pi$. 
At $\theta=0$,  Yang-Mills theory is believed to possess a unique vacuum. If so, at $\theta=\pi$, the theory must have two vacua, and spontaneously broken CP. 

In order to see how this symmetry is realized in the long distance theory, recall the 
two types of monopole amplitudes  (\ref{2amplitudes}), ${\cal M}_1$ and  ${\cal M}_2$. These amplitudes are periodic functions of $\sigma \in [0, 2\pi)$, leading to the Lagrangian (\ref{lag1}). 
Since the microscopic theory possess an exact  $\Z_2$ symmetry exactly at (odd-multiples of)  $\theta=\pi$, and no other $\theta$,  this  must also be a symmetry of the  low-energy effective theory at exactly at (odd-multiples of)  $\theta=\pi$, and no other $\theta$.

 Consider the shift $\sigma \rightarrow \sigma + \psi$. This rotates  the amplitudes as 
 \begin{equation}
 {\cal M}_1  \rightarrow e^{i \psi  } {\cal M}_1, \qquad {\cal M}_2\rightarrow e^{-i \psi  } {\cal M}_2, 
\qquad  [{\cal M}_1  \overline {\cal M}_2]  \rightarrow e^{ 2i \psi  } [{\cal M}_1  \overline {\cal M}_2] \, . 
\end{equation}
Clearly, this is not a symmetry of   (\ref{pot})  for general $\psi$.  
However, only at  $\psi=\pi$,   the phase shift of both monopole amplitudes  coincide ${\cal M}_i  \rightarrow (-1) {\cal M}_i$,  and bion amplitude remains invariant. Consequently, in low energy effective theory (\ref{pot}),  
$\cos \textstyle \left(\frac{\theta}{2} \right)   \cos \sigma   \rightarrow - \cos \textstyle \left(\frac{\theta}{2} \right)   \cos \sigma$ and  $ \cos 2 \sigma   \rightarrow  \cos 2 \sigma$. This can be a symmetry of the theory if and only if the first operator vanishes identically. This happens exactly at (odd-multiples of)  $\theta=\pi$. 

The low-energy effective theory has a $\Z_2$ shift symmetry exactly at $\theta=\pi$, and is 
 described by the  Lagrangian
\begin{eqnarray}
L^{\rm d} =&& \frac{1}{2L} \left( \frac{g}{4\pi} \right)^2  (\nabla  \sigma)^2  - 2  b  e^{-2S_0}  
\cos 2  \sigma   + O(e^{-4S_0} \cos 4\sigma  )
   \end{eqnarray}
   The effective theory obtained in  deformed Yang-Mills theory  at $\theta=\pi$ coincides with the one in non-linear sigma models  \cite{Senthil}.
   The potential has two minima within the unit cell related by the $\Z_2$ shift-symmetry  $\sigma \rightarrow \sigma+ \pi$,    and a spontaneously broken CP-symmetry. 
 CP, in the small- $S^1$ domain, is broken due to the condensation of a disorder (monopole) operator, 
\begin{equation} 
e^{-S_0} \langle e^{i \sigma} \rangle = \pm  e^{-S_0} 
\end{equation}
Due to spontaneous breaking of CP, there must be a domain wall. Consider one filling  $\R^2$ on $xy$ plane. Then, the $\sigma(z)$ must interpolate between the two vacua such that
 $   \int_{-\infty}^{\infty} d \sigma =  \pi\,$.
The resulting domain wall tension scales as 
  $ T_{\rm DW} (\pi) \sim   \Lambda^3 (\Lambda L)^{2/3}$.

Clearly, as the $\theta$ parameter is varied, there are not only quantitative but qualitative changes  in the behavior of gauge theory.  At $\theta=\pi$, despite the fact that  the density of monopole-instantons is exponentially larger than the density of magnetic  bions,  confinement, the mass gap, and string tensions are sourced by the latter, and  the theory  has two  vacua.

\subsection{Continuity and evading  the problems with  4d instantons}
The problems associated with 4d instantons  in an unbroken asymptotically free gauge theory  on $\R^4$ are  well-know. Since the instanton size is a moduli, a self-consistent treatment of  dilute instanton gas approximation does not exist. (See, for example, section 3.6 in Coleman's lecture  \cite{Coleman}. This is still an  up to date  presentation.)

    In the semi-classical regime,  the deformed theory   exhibits abelianization, and the long distance theory is described by $ SU(2) \rightarrow U(1)$ abelian group, much like the Coulomb branch of supersymmetric theories. The gauge symmetry breaking scale is $v \sim \frac{1}{L}$. 
In our  locally four-dimensional spontaneously broken gauge theory, the instanton  size moduli   is cut-off by the gauge symmetry breaking scale  $v$, as in supersymmetric gauge theories with 
adjoint scalars, such as $\N=4$ SYM.    This sets the scale of the   coupling constant entering to the 4d instanton amplitude $\exp\left[ -\frac{8 \pi^2 }{g^2(v)} + i \theta \right]$.  The only   4d instantons in the systems are the ones with size less than $v^{-1} \sim L$.  Therefore, the 4d instanton expansion is justified.  

 However, as discussed in depth, the  control over the 4d instantons is hardly the point.
   The expansion on 
 $\R^3 \times  S^1$ is an expansion in monopole-instantons. It is the 3d instantons and  twisted-instantons   (whose topological charge in a center-symmetric background is  $1/2$). 
 For general $N$, the topological charge for these defects is 
  $1/N$, and the correct expansion parameter is 
 \begin{equation}
 \label{FI}
 \exp\left[-\frac{8 \pi^2 }{g^2N(v)} + i \frac{\theta}{N}\right]  
 \end{equation}
In the semi-classical  expansion,  the   4d instantons with amplitude  $ \sim \exp[
 -\frac{8 \pi^2 }{g^2}]$ are  exponentially  suppressed and are not the origin of the most interesting physics. The expansion parameter is  (\ref{FI}), and not the 4d instanton amplitude. It is worth noting that  (\ref{FI}) survives the large-$N$ limit.

\section{Quantum anti-ferromagnets and deformed Yang-Mills}
In this section, we will outline a surprising  relation  between two dimensional quantum anti-ferromagnets (AF) on bi-partite lattices,  deformed Yang-Mills theory on 
$\R^{3} \times S^1$, and by continuity, pure Yang-Mills theory on $\R^4$. 
 As reviewed below, the long distance theory for the AF is defined on 
$\R^{2,1}$ in Minkowski space and the one of the dYM is also defined on $\R^{2,1}$. We will 
demonstrate that AF with even and odd integer spin (not half-integer) is equivalent to dYM with 
$\theta=0$ and  $\theta=\pi$, respectively. 

  The ground state properties of $SU(N)$  quantum anti-ferromagnets on  bi-partite lattices in    two spatial dimensions  are  studied in  Ref.~\cite{Read:1990zza}. Following Ref.~\cite{Read:1990zza}, call the two-sublattices of the bi-partite lattice as $A$ and $B$. One associates an irreducible representation of $SU(N)$ with $n_c$ rows and $m$-columns to sublattice-$A$ and the conjugate irrep  with $n_c$ rows and $N-m$-columns to 
sublattice-$B$. 
   For $SU(N)$, in the low energy,  large $n_c$ (spin) limit,  the continuum limit of the lattice system    can be described by a NL$\sigma$ model 
with  a complex Grassmann manifold  (target space)
\begin{equation}
M_{N,m}(\mathbb C)= U(N) / \left[ U(m) \times U(N-m) \right] 
\end{equation}   
     supplemented with a Berry phase induced term. 
   For $m=1$, this corresponds to the
      $\mathbb C\mathbb P^{N-1}$ model.     
     The  field theory has topological configurations, ``hedgehog" 
  type instanton events. Ref.~\cite{Read:1990zza} expresses 
   the low energy partition function as a dilute gas of  instantons with complex fugacities.
   The complexification of the fugacity is due to  the  Berry phase.   Ref.\cite{Read:1990zza}  proposed that the properties of the Coulomb plasma vary  periodically with the spin $n_c$ of states on each site, and that the ground state has a degeneracy 
  \begin{equation} 
  d(2S) = 1, 4,2, 4, \qquad  
   {\rm for}  \;\;\;\;  n_c= 2S= 0,1,2,3 \; {\rm  (mod 4)}  
  \end{equation} 
According to Ref.~\cite{Read:1990zza},  for a given  $n_c$,  the fugacity of the monopole-instantons becomes complex due to the Berry phase. The monopole amplitude is modified into  
\begin{equation}
 e^{ - S_0 }  e^{ i \widetilde \sigma}   \longrightarrow e^{ - S_0 +  i \frac{ \pi n_c}{2} \zeta_s }  e^{ i  \widetilde \sigma}, \qquad  \zeta_s=0,1,2,3
\end{equation}
Since the lattice is bi-partite, the unit cell of the lattice, similarly to  staggered fermions in lattice gauge theory, may be thought of as having a unit cell  $2\mathfrak{a} \times  2\mathfrak{a}$. The monopole-events emanating from each one of these four smaller cells 
(with size $\mathfrak{a} \times  \mathfrak{a}$)   may acquire a different phase depending on the value of  $n_c$.  There are three inequivalent  cases.

{\bf i)} For $n_c=0$ (mod 4), the phase is zero. Then,  there is  only one type of monopole-instanton event  
 \begin{equation}
 {\cal M}_1 \sim  e^{ - S_0 }  e^{ i  \widetilde \sigma},
 \end{equation}
 whose proliferation generates the effective potential $V(n_c=0) \sim e^{-S_0}  
\cos  \widetilde \sigma  $  with a unique ground state. 
 
{\bf ii)}  For $n_c=2$ (mod 4), then  there are two types of instanton events, which differ by a phase shift $\pi$: 
 \begin{equation}
 {\cal M}_1 \sim  e^{ - S_0 }  e^{ i \widetilde \sigma}, \qquad {\cal M}_2 \sim  e^{ - S_0 + i \pi   } e^{ i  \widetilde \sigma }
 \label{2ev}
 \end{equation}
  Clearly, these two events, in a Euclidean path integral formulation, interfere destructively, and the effective potential is $
V(n_c=2) \sim e^{-2S_0}  
\cos 2  \widetilde \sigma  $   with two ground states. 
 
 {\bf iii)}  For $n_c=1,3$ (mod 4), then  there are four types of instanton events, 
  \begin{equation}
  {\cal M}_1 \sim  e^{ - S_0 }  e^{ i  \widetilde \sigma}, \;\; \; {\cal M}_2 \sim  e^{ - S_0 +  i  \frac{\pi}{2} } e^{ i  \widetilde \sigma}, \;\;\;  {\cal M}_3 \sim  e^{ - S_0 +  i  \pi  } e^{ i  \widetilde \sigma}, \;\;\;  {\cal M}_4 \sim  e^{ - S_0 +  i  \frac{3\pi}{2} } e^{ i  \widetilde \sigma}
  \end{equation} 
  These instanton events interfere destructively  both at leading order $(e^{-S_0})$, as well as subleading   orders   $(e^{-2S_0},e^{-3S_0})$. The effective potential is 
$  V(n_c=1) \sim e^{-4S_0}  
\cos 4  \widetilde \sigma  $
  with four ground states.

Now, let us switch back to deformed Yang-Mills theory. This theory has two types of monopoles, 
$  {\cal M}_1 $ and  $ {\cal M}_2 $. At $\theta=0$, the amplitude $  {\cal M}_1 $ and  
$ \overline {\cal M}_2 $ are identical. The theory  at $\theta=0$  (mod $2\pi$) has a unique ground state, much like the $n_c=0$ (mod 4) case of the spin system. 
However, when  we introduce $\theta$, we can in fact distinguish  $  {\cal M}_1 $ and  
$ \overline {\cal M}_2 $ monopole-events.They have identical magnetic charge, but  their topological phase are opposite in sign.

 Using (\ref{lag2mod}), the grand canonical partition function of the Coulomb plasma
  takes the form 
\begin{equation}
Z(\theta)= \sum_{\substack{ n_1, \n_1 \geq 0  \\ \n_2, \n_2 \geq 0 } }  
\; \sum_{ n_{\rm b}, \n_{\rm b} \geq 0 }
e^{i  \theta \left[(n_2 - \n_2) + ( n_{\rm b}-  \n_{\rm b}  )   \right] }   
\; Z({n_1 n_2 \n_1 \n_2, n_{\rm b} \n_{\rm b}} )
\label{partition}
\end{equation}
where $   Z({n_1 n_2 \n_1 \n_2, n_{\rm b} \n_{\rm b} }) $ is the canonical partition function for a fixed number of monopole-instantons, bions. 
  The crucial difference with respect to Polyakov model  --- apart from the existence of  ${\cal M}_{2}$ monopole --- is  the existence of the  $\theta$-phase factor. The partition function is 
  $2 \pi $ periodic.
    

 The partition functions of  spin system  with integer spin, for the first two cases listed above, 
  are  
\begin{eqnarray}
&& S \in 2\Z   \qquad  \;\;\;\; \;\; \Longrightarrow   \qquad   Z= \sum_{n_1, n_2, \n_1, \n_2 \geq 0}   Z_{n_1 n_2 \n_1 \n_2} \cr
&&  S \in 2\Z+1   \qquad \Longrightarrow   \qquad   Z= \sum_{n_1, n_2, \n_1, \n_2 \geq 0} e^{i  \pi[(n_2 - \n_2)  + ( n_{\rm b}-  \n_{\rm b}  )] } Z({n_1 n_2 \n_1 \n_2 , n_{\rm b} \n_{\rm b} })
\end{eqnarray} 
which means that the deformed YM theory
interpolates  between even integer spin $S \in 2\Z$  and odd-integer spin $S \in 2\Z +1$ as $\theta$ varies continuously from $0$ to $\pi$,.   
In the  $S \in 2\Z $ partition function, we did not include bions because they give an  exponentially suppressed  perturbation.

We reach to the following identification between the quantum anti-ferromagnet with spin $S$ and 
deformed YM theory with $\theta$ angle: 
\begin{eqnarray}
{\rm dYM \;\; at \;\;}   \theta=0 \; \;  ({\rm mod} \; 2 \pi)  \qquad \Longleftrightarrow  \qquad 
{\rm AF \;\; at \;\;}   2S=0 \; \;  ({\rm mod} \; 4)  \cr 
 {\rm dYM \;\; at \;\;}   \theta=\pi \; \;  ({\rm mod} \; 2 \pi)  \qquad \Longleftrightarrow  \qquad 
{\rm AF \;\; at \;\;}   2S=2 \; \;  ({\rm mod} \; 4)  
\end{eqnarray} 
Spin in the AF  is discrete, whereas the $\theta$ angle is continuous. Nonetheless, by  inspecting 
(\ref{partition}), we may identify\footnote{The identification for the one dimensional spin chain (1+1 dimensional field theory) would be 
 $\theta \Leftrightarrow 2\pi S $, and in that case, the difference is between the integer and half-integer spin. Gauge theory, however, is related to spin systems in two spatial dimensions. }
\begin{eqnarray}
\theta \Longleftrightarrow \pi S 
\end{eqnarray}
There is a sense in which the $\theta$ angle in YM theory may be seen as a continuous  version of the  discrete spin variable in the quantum spin system. The topological phase in Yang-Mills theory can be identified with the Berry phase induced topological term in  the 
$M_{N,m}(\mathbb C)$ NL$\sigma$-model.

Note that the deformed YM theory  does not  capture the half-integer spin cases. 
 For that, one needs four different types of monopole instanton events, while dYM has only two types.

\subsection{Berry phase versus 4d topological phase}
\label{berrytop}
It may sound surprising that Berry phase in the AF spin-system and topological phase in 4d gauge theory may actually be identified. Both systems, in their long distance descriptions,  can be formulated on $\R^3$ in a Euclidean space. 

However, it is well-known on $\R^3$ that an analog of the topological term of the 4d theory does not exists. There is a 3d Chern-Simons term, but that does not play a role in our problem; in fact, it would have  been detrimental for the survival of long-range interactions between monopoles.  Then, it is crucial  to understand, from a {\it 3d long distance point of view}, 
 how  the compactified theory generates a topological phase for monopole-instantons. 
This helps us to see why the effect of Berry phase induced action and the effect of the topological phase are actually the same thing.

 Ref.~\cite{Read:1990zza}  shows, in some detail,  that in the long-distance description of the 
 quantum anti-ferromagnets on  bi-partite lattice, there exist a   Berry phase induce term in 
 the  effective action given by 
 \begin{equation}
 S_B= \sum_s i  \frac{n_c \pi}{2}\zeta_s  \times m_s \qquad  
 m_s =  \frac{1}{4\pi} \int_{S^2_\infty} B.dS =  \frac{1}{4\pi} \int_{\R^3} \nabla  B  \;.
\label{Berry2}
 \end{equation}   We will not repeat their derivation here, and  refer the reader to 
 Ref.~\cite{Read:1990zza}  for details. 

The topological term in the locally four-dimensional Yang-Mills action, formulated on 
$ \R^3 \times S^1$,  is the second Chern number.  How does it relate to Berry phase induced term $S_B$, and more specifically, how does the  first Chern number, the magnetic flux, even appear in the long-distance description?  Below, we will demonstrate  the following statement connecting the two.
\begin{quote}
The second Chern-number on $\R^4$, upon compactification on   $\R^3 \times S^1$ and in a background of a center-symmetric  gauge holonomy, gives a contribution proportional to first Chern-number (magnetic flux) of the topological configuration times $(\pm \half)$ depending on the type of the topological defect.  In other words, the center-symmetric `dimensional reduction' of the 4d topological $\theta$ term is the Berry phase induced action (\ref{Berry2}) in anti-ferromagnets.
\end{quote}
The steps necessary to demonstrate this statement are already present in my  work with 
Poppitz in Ref.~\cite{Poppitz:2008hr} on index theorem on  $\R^3 \times S^1$.
Consider the topological charge contribution  in the action.
 \begin{equation}
 \label{indexstatic1}
Q ={1 \over 16 \pi^2} \int_{\R^3 \times S^1}   \;  \tr F_{\mu\nu} \tilde{F}_{\mu\nu} 
= {1 \over 32\pi^2} \int_{\R^3 \times S^1} \partial_\mu  K^\mu~,
\end{equation}
The topological charge density is a total derivative and can be written as the divergence of the 
topological current  $K^\mu$:
 \begin{equation}
 \label{topocurrent}
 K^\mu = 4 \epsilon^{\mu\nu\lambda\kappa} \tr \left( A_\nu \partial_\lambda A_\kappa + { 2 i \over 3} \; A_\nu A_\lambda A_\kappa \right)~.
 \end{equation}
 Consider the ${\cal M}_1$ monopole.   Using the fact that for   the static BPS background  $K^\mu$ is  a periodic function of the compact coordinate $y$, we may re-write
$$\int_{\R^3 \times S^1}  \partial_\mu  K^\mu =  \int d^3 x \int_0^L  dy  \left(
  \partial_4 K_4 +    \partial_m K_m\right) = L \int_{\R^3} \;  \partial_m K_m \; .$$
   $K_m$ is the  spatial component of $K^\mu$, given by
 \begin{equation}
K^m = 4 \epsilon^{mij} \tr \left(   A_4 F_{ij} -   A_i \partial_4 A_j  - \partial_i (A_4 A_j)\right)~.
\end{equation}
The only contribution to topological charges comes from the first term, which, 
using $\epsilon^{ijk}F_{jk} = 2 B^i$, can be written as 
$ 8 \tr   A_4 B_m $. This is  the gauge invariant magnetic field in the dimensionally reduced theory. This means that we can replace the spatial component of the topological current with the magnetic field under the integral sign, namely  $\int K_m=  \int 4 v  B_m$. 
Using the explicit form of the gauge holonomy and the asymptotic form of the magnetic field, we obtain $
 8 \tr   A_4 B_m\big\vert_\infty  =     \frac{4\pi}{L} {\hat{r}^m \over r^2}~.$
Thus,  the topological charge contribution reduces to 
 \begin{equation}
 \label{indexstatic2}
Q ({\cal M}_1) = \frac{1}{2} \frac{1}{4 \pi} \int_{\R^3}  \nabla B  =  \frac{1}{2}  \frac{1}{4 \pi}  \int_{S^2_{\infty}} B.dS = +\frac{1}{2}  
\end{equation}
Similar calculation for the $\overline {\cal M}_2$ anti-monopole (or twisted anti-monopole) is more technical due to twist.  The  magnetic charge  of $\overline {\cal M}_2$ is also   $+1$. 
Using the result of Section 2.2 of Ref.~\cite{Poppitz:2008hr}, we find 
the phase associated with  $\overline {\cal M}_2$-event as 
\begin{equation}
 \label{indexstatic3}
Q (\overline {\cal M}_2) = - \frac{1}{2} \frac{1}{4 \pi} \int_{\R^3}  \nabla B  =  - \frac{1}{2}  \frac{1}{4 \pi}  \int_{S^2_{\infty}} B.dS = -\frac{1}{2}  
\end{equation}
As noted in (\ref{2amplitudes}), despite the fact that 
${\cal M}_1$  and $\overline {\cal M}_2$ have the same magnetic charge, they acquire opposite topological phases upon introducing the $\theta$ angle.  We obtain
\begin{eqnarray}
\label{magical}
\exp\left[ {{ i \theta  \over 32 \pi^2} \int_{\R^3 \times S^1}    F_{\mu\nu}^a \tilde{F}_{\mu\nu}^a  }  \right]  && =
\exp \left[{  \pm i \frac{\theta}{2}  \frac{1}{4\pi} \int_{\R^3}  \nabla B } \right]  \cr
&&= 
\exp \left[{ \pm i \frac{\theta}{2}  \frac{1}{4\pi}  \int_{S^2_{\infty}} B.dS } \right]  =\exp \left[{ \pm i \frac{\theta}{2}   }\right] \;,   \qquad \qquad
\end{eqnarray}
respectively, for ${\cal M}_1$ (+)  and $\overline {\cal M}_2$ $(-)$. 
This   relation underlies  the topological interference effects. 
It is also the reason why the topological phase in gauge theory on 
$\R^3 \times S^1$ and Berry phase induced action in  quantum anti-ferromagnets on $\R^{2,1}$ 
($\R^3$ in Euclidean formulation)   coincides for certain values of $\theta$, and that the phenomena that we have uncovered are a generalization of the physics of Berry phases of spin systems.

Eq.~(\ref{magical}) also instructs us that the sign problem in simulations of 
quantum anti-ferromagnets and Yang-Mills theory with $\theta$ angle are equivalent problems in their respective semi-classical regimes. 

\section{Discussion and prospects}
 
As an end note, we  would like to mention few ways to generalize this work and a new problem 
in gauge theory.

{\it Generalization:} Deformations and continuity can be used to generalize our work to all gauge groups. 
A  more accessible theory  is $SU(N)$ QCD(adj) with light fermions  endowed with periodic 
(not anti-periodic)  boundary conditions. This theory automatically satisfies our continuity criterion. Moreover,  by dialing the fermion mass term, it can be continuously connected to 
Yang-Mills theory. 

{\it  Mapping field theory $\theta$-angle to Aharonov-Bohm effect:} One direction that we  find interesting is a  more direct link between the Aharonov-Bohm effect in  ordinary quantum mechanics and $SU(N)$ gauge theory with $\theta$ angle. A certain modification of the $T_N(\theta)$  model   is related to  quantum field theory by using  compactification on asymmetric three-torus. On torus,  the study of zero mode dynamics and magnetic  flux sectors  reduce to  a basic quantum mechanics problem with an Aharonov-Bohm flux  \cite{Poppitz}.  Mapping the   $\theta$ angle dependence of Yang-Mills theory (in a  semi-classical domain)   to  Aharonov-Bohm effect,  the effects of a changing  $\theta$ 
and CP-symmetry breaking can be emulated through (superselection sectors) in quantum mechanics. 

{\it  What is the $\theta$-angle in 4d gauge theory?}  Our construction also suggests that the $\theta$ parameter  of Yang-Mills theory 
  may have a more  interesting  topological interpretation. 
  Recall the  topological terms in 4d gauge theory  and  in quantum mechanics of a charged particle on a circle, 
\begin{equation}  
  { i \theta  \over 16 \pi^2} \int   \; \tr F_{\mu\nu} \tilde{F}_{\mu\nu} , \;\; \;\;  {\rm and} \;\; \;\;  \frac{i \theta^{\rm qm}}{2\pi} \int \dot q
  \end{equation}
   In quantum mechanics,   
the presence of the theta term   be reformulated as a ``hole" in  the  topology of the configuration space $q(t)$, and 
\begin{equation}
{\theta}^{\rm qm} \equiv  \frac{ |e| \Phi}{ \hbar c} = \frac{ |e|}{ \hbar c} {\int \vec B^{\rm em} d\vec S} =  \frac{ |e|}{ \hbar c} {\int \vec A^{\rm em} d\vec l} 
\end{equation}
where $B^{\rm em}$ and $A^{\rm em}$ are  the magnetic field and gauge potential of  electromagnetism. This term follows from the usual minimal coupling, 
$e  \vec q. \vec A^{\rm em} $. We can re-write the topological term in 
quantum mechanics as 
\begin{equation}  
 \frac{i}{2\pi}   \theta^{\rm qm} \int \dot q =  
\frac{i }{2\pi} \left( \frac{ |e|}{ \hbar c} {\int \vec B^{\rm em} d\vec S}   \right)   \times  \int \dot q
\label{curious}
  \end{equation}
We can see the tiny solenoid  which supports the  $\vec B^{\rm em}$ flux as  drilling  a hole in the configuration space  and turning it a non-simply connected space. This gives $\theta$ angle a physical  meaning in quantum mechanics.

The question we are curious about is the analog of the (\ref{curious}) in quantum field theory. 
Perhaps, $\theta$ angle in Yang-Mills can  be reformulated as a ``hole" in  the  topology of the configuration space $A(\vec x)$,  much like the Aharonov-Bohm effect.
 It would be interesting to understand the change in the topology of the configuration space of gauge theory   which would induce the  4d  $\theta$ term. 
 At another layer of abstraction, it would also be useful to understand   the origin of the 
 $\theta$-``flux" in gauge theory.



  \acknowledgments 
  I  thank  Philip Argyres, Adi Armoni, Aleksey Cherman, Gerald Dunne, Gregory Gabadadze, 
  Leonardo Giusti, Dima Kharzeev,  Mehmet \"Ozg\"ur Oktel and  Erich Poppitz   for  discussions on various topics relevant to this paper.

\end{document}